\documentclass[useAMS,usenatbib,usegraphicx]{mn2e}
\usepackage{psfig,amssymb}
\usepackage{times,color}
\usepackage{amsmath}
\newcommand{\beq}{\begin{equation}}
\newcommand{\eeq}{\end{equation}}

\begin{document}
\title[]{$\gamma$-rays from annihilating dark matter in galaxy
  clusters: stacking {\em vs} single source analysis}
\author[Nezri, White, Combet et al.]{E. Nezri$^{1}$\thanks{E-mails:Emmanuel.Nezri@oamp.fr (EN),
richard.white@leicester.ac.uk (RW),
celine.combet@lpsc.in2p2.fr (CC), jah85@leicester.ac.uk (JAH), dmaurin@lpsc.in2p3.fr (DM),
etienne.pointecouteau@irap.omp.eu (EP)},
        R. White$^{2}$\footnotemark[1],
        C. Combet$^{3}$\footnotemark[1],
        J.A. Hinton$^{2}$\footnotemark[1],
        D. Maurin$^{3}$\footnotemark[1],
        E. Pointecouteau$^{4,5}$\footnotemark[1]\\
  $^1$Laboratoire d'Astrophysique de Marseille - LAM, Universit\'e d'Aix-Marseille \& CNRS, UMR7326, 38 rue F. Joliot-Curie, 13388 Marseille Cedex 13, France\\
  $^2$Dept. of Physics and Astronomy, University of Leicester, Leicester, LE1 7RH, UK\\
  $^3$Laboratoire de Physique Subatomique et de Cosmologie,
      Universit\'e Joseph Fourier Grenoble 1/CNRS/IN2P3/INPG,
      53 avenue des Martyrs, 38026 Grenoble, France\\
 $^4$Universit\'e de Toulouse (UPS-OMP), Institut de Recherche en Astrophysique et Plan\'etologie\\
  $^5$CNRS, UMR 5277, 9 Av. colonel Roche, BP 44346, F 31028 Toulouse cedex 4, France
}
\pagerange{\pageref{firstpage}--\pageref{lastpage}} \pubyear{Xxxx}
\date{Accepted Xxxx. Received Xxxx; in original form Xxxx}
\label{firstpage}

\maketitle

\begin{abstract}
 
  Clusters of galaxies are potentially important targets for indirect
  searches for dark matter annihilation. Here we reassess the
  detection prospects for annihilation in massive halos, based on a
  statistical investigation of 1743 clusters in the new Meta-Catalog
  of X-ray Clusters. We derive a new limit for the extra-galactic dark
  matter annihilation background of at least 20\% of that originating
  from the Galaxy for an integration angle of $0.1^\circ$. The number
  of clusters scales as a power law with
  their brightness (boosted by dark-matter substructures), suggesting
  that stacking may provide a significant improvement over a single
  target analysis. The mean
  angle containing $80\%$ of the dark-matter signal for the sample
  (assuming an NFW DM profile) is $\sim 0.15^\circ$ (excluding the
  contribution from the PSF of any instrument), indicating that
  instruments with this angular resolution or better would be optimal
  for a cluster annihilation search based on stacking.
  A detailed study based on the Fermi-LAT performance
  and position-dependent background, suggests that stacking may result
  in a factor $\sim$2 improvement in sensitivity, depending on the
  source selection criteria. Based on the expected performance of CTA,
  we find no improvement with stacking, due to the requirement for
  pointed observations. We note that several potentially important
  targets: Opiuchius, A\,2199, A\,3627 (Norma) and
  CIZA\,J1324.7$-$5736 may be disfavoured due to a poor contrast with
  respect to the Galactic dark-matter signal. 
  The use of the homogenised MCXC meta-catalogue provides a robust ranking
  of the targets, although the absolute value of their signal
  depends on the exact dark matter substructure content. For conservative
  assumptions, we find that galaxy clusters (with or without stacking) can
  probe $\langle\sigma v\rangle$ down to $10^{-25}-10^{-24}$~cm$^3$~s$^{-1}$
  for dark matter masses in the range 10 GeV - 100 GeV. For more favourable 
  substructure configurations, $\langle\sigma v\rangle\sim 10^{-26}$~cm$^3$~s$^{-1}$
  may be reached.

\end{abstract}


\section{Introduction}
\label{sec:intro}

The annihilation of dark matter (DM) particles into $\gamma-$rays has been
flagged as one of the most promising channels for indirect detection. Regions of
high DM density are of particular interest, making the Galactic centre the most
obvious target \citep{1987ApJ...313L..47S}. However, the Galactic centre is
plagued by a large astrophysical $\gamma-$ray background at all angular scales
that makes any DM signal difficult to identify
\citep[e.g.,][]{2004A&A...425L..13A}. In that respect, dwarf spheroidal galaxies
(dSphs) have the advantage to be essentially background-free, relatively close
by and with DM density profiles that can be constrained from their internal
kinematics. This has made them popular candidates for indirect detection
\citep{Evans:2003sc,2006PhRvD..73f3510B,2007PhRvD..75h3526S,2009MNRAS.399.2033P,
abdo10,2010AdAst2010E..45K,2011JCAP...12..011S,2011ApJ...733L..46W,2011MNRAS.418.1526C,2011PhRvL.107x1302A}. 

Somewhat less explored to date, clusters of galaxies are the largest
gravitationally bound structures in the universe, the large DM content
of which makes them potentially interesting targets for indirect
detection \citep{2006A&A...455...21C}. Although strong constraints
have already been derived from X-ray and gravitational lensing studies
on the DM distribution in clusters \citep{2005A&A...435....1P,vikhlinin06,2007ApJ...664..123B,2010MNRAS.406.1134S,
  2011A&A...531A.169P,ettori11}, constraining the inner DM
distribution is still a challenging task. Even strong lensing, which likely
is the best suited way to pin down the DM distribution at the
cluster centre, fails to assemble convincing constraints (see for
instance the different conclusions reached by
\citealt{2007ApJ...668..643L,2011ApJ...728L..39N,2012MNRAS.421.3147M}).
Estimates of the DM profile and calculations of the $\gamma-$ray flux
from clusters are based on X-ray observations, from which NFW
\citep{navarro97} or Einasto \citep[e.g.,][]{2006AJ....132.2685M} profiles are
assumed. For instance, based on the HIFLUGCS catalogue containing 106
objects \citep{2002ApJ...567..716R,2007A&A...466..805C}, several
authors have identified the potentially most luminous objects in DM
emission, such as Fornax, Coma or Perseus \citep{2009PhRvD..80b3005J,
  2011PhRvD..84l3509P}.  The non-detection of these favoured targets
by Fermi-LAT and H.E.S.S. has resulted in constraints on the DM
annihilation cross-section
\citep{2010JCAP...05..025A,2010PhRvD..82b3506Y,2012arXiv1201.0753A,2012ApJ...750..123A}. See,
however, \citet{2012arXiv1201.1003H} for a possible evidence of an
extended emission. Alternatively to these `observational' approaches,
\citet{2011ApJ...726L...6C} have performed synthetic Fermi
observations from the CLUES constrained cosmological N-body simulation
of the local universe and flagged Virgo and Coma, along with DM
filaments as interesting targets.  \citet{2012MNRAS.419.1721G} is
another example of high-resolution N-body simulations used to estimate
the DM profile/content and signal of selected targets.

In this study we make use of the recently published Meta-Catalogue of X-ray
Clusters, MCXC \citep{piffaretti11}, which contains 1743 clusters of galaxies.
The size of the catalogue, with $\sim 17$ times more objects than the HIFLUGCS
catalogue, makes it possible to investigate some statistical aspects of DM
indirect detection in galaxy clusters. This paper is part of a series: a first
paper \citep{2012PhRvD..85f3517C} highlighted the improvement brought by a
stacking analysis over a single source analysis for the DM decay case. The
current paper focuses on the DM annihilation case: we provide a quantitative
analysis of the best observing strategy to use for the Fermi-LAT and CTA
observatories, we discuss the potential benefit of a stacking strategy with 
respect to single source observation, and we also present the number of objects
to look at to optimise detectability. The last paper of the series addresses the
possibility of using the stacking analysis to disentangle CR-induced from
DM-induced signal \citep{2012arXiv1203.1166M}.

The paper is organised as follows: in Section~\ref{sec:model}, we briefly present
the key quantities for the signal calculation ($J$-factor, DM halo profiles). In
Section~\ref{sec:clusterhalosample}, the MCXC catalogue is introduced, and the cluster signal
distribution presented, along with the resulting skymap.  The contrast with the
Galactic DM annihilation signal and the astrophysical background, and the
consequences for the ranking of the best targets are also discussed. The stacking
approach and results are presented in Section~\ref{sec:stacking}. In
particular, the boost of the DM signal from DM substructures (in the galaxy clusters) and its
effect on the stacking is detailed. The sensitivity to a DM signal taking into
account realistic instrumental responses is then evaluated for Fermi-LAT and CTA
instruments. We conclude in Section~\ref{sec:discussion}. (Appendix~\ref{app1}
provides parametric formulae to evaluate the signal from a cluster for any
integration angle. Appendix~\ref{app2} provides a quick comparison to values of $J$
found in other works).

\section{The model and its ingredients}
\label{sec:model}

The $\gamma$-ray flux $\Phi_{\gamma}$ from dark matter annihilations
(cm$^{-2}$~s$^{-1}$~sr$^{-1}$~GeV$^{-1})$ received on Earth in a solid angle
$\Delta\Omega$, is given by\footnote{We remind that the spatial term $J$ in
Eq.~(\ref{eq:flux}) couples to the energy-dependent term
$dN_{\gamma}/dE_{\gamma}$ for objects at cosmological distances, because
$\gamma$-rays are absorbed along the line of sight (e.g.,
\citealt{2010NuPhB.840..284C}). The redshift distribution of the MCXC
catalogue of galaxy clusters \citep{piffaretti11} peaks at $z\sim 0.1$ (see their
Fig.~1): following \citet{2012PhRvD..85f3517C}, we neglect the absorption
for the MCXC galaxy clusters.}
\begin{equation}
     \frac{d\Phi_{\gamma}}{dE_{\gamma}}
        = \frac{1}{4\pi}\frac{\langle\sigma_{\rm ann}v\rangle}{\delta m_{\chi}^{2}}
          \cdot \frac{dN_{\gamma}}{dE_{\gamma}} \times J(\Delta\Omega),
\label{eq:flux}
\end{equation}
where $\delta=2$ for a self-conjugate particle and 4 otherwise, $m_{\chi}$ is the particle mass, $\langle \sigma_{\rm ann}v\rangle$ is the
velocity-averaged annihilation cross section, $dN_{\gamma}/dE_{\gamma}$ is the
energy spectrum of annihilation products. 

\subsection{Spectrum and astrophysical factor $J$}

The differential annihilation spectrum, $dN_{\gamma}/dE_{\gamma}$,
requires a specific DM particle model. It is the sum of a prompt
contribution and a contribution from inverse Compton scattered (ICS)
secondary electrons and positrons with the CMB \citep[see,
e.g.,][]{2012JCAP...01..042H}. For the sake of simplicity and to keep
the analysis as DM particle model-independent as possible, we
disregard the `delayed' ICS contribution. The latter has a similar
spatial distribution to that of the prompt
\citep{2012JCAP...01..042H}, so that the factorisation of the spatial
and energy-dependent term in Eq.~(\ref{eq:flux}) holds. Actually,
depending on the annihilation channel, the ICS contribution can
dominate over the prompt one.  Considering only the prompt
contribution as we do here provides a conservative and robust lower
limit on detectability. In this paper, we further restrict ourselves
to the $b\bar{b}$ annihilation channel, taken from Eq.~(6) and Table
XXII in \citet{2011PhRvD..83h3507C}. We note that the spectral
parameters in \citet{2011PhRvD..83h3507C} are provided for WIMP
masses in the range of 50 GeV to 8 TeV. Here we assume the spectral
parameters for masses below 50 GeV are given by the parameters for a
50 GeV mass, and similarly above 8 TeV. The results are not strongly
affected (less than a factor 1.5 in the sensitivity limits) by 
 the choice of the $\gamma$-ray annihilation channel (apart from the
$\tau\bar{\tau}$ channel).

The `$J$-factor' represents the astrophysical contribution to the signal and corresponds 
to the integral of the squared dark matter density, $\rho^2(l,\Omega)$,
over line of sight $l$ and solid angle $\Delta\Omega$,
\beq
J(\Delta\Omega)=\int_{\Delta\Omega}\int \rho^2 (l,\Omega) \,dld\Omega\;.
\label{eq:J}
\eeq 
We have $\Delta\Omega = 2\pi\cdot(1-\cos(\alpha_{\rm int}))$, and $\alpha_{\rm
int}$ is referred to as the `integration angle' in the following.  All $J$-factors
presented below, including substructures (in the Galaxy or in galaxy clusters), are
calculated from the public code {\sc clumpy} v2011.09 \citep{2012CoPhC.183..656C}.

\subsection{The smooth DM halo and substructures}
\label{sec:ingred_DM}

For the DM halo smooth profile, we use an NFW \citep{navarro97} 
\begin{equation} 
  \displaystyle
  \rho(r)=\frac{\rho_s}{\left(\frac{r}{r_s}\right)\left(1+\frac{r}{r_s}\right)^2}\,,
  \label{eq:rhoNFW}
\end{equation}
where $r_s$ is the scale radius and $\rho_s$ is the normalisation\footnote{A
decreasing inner slope with the halo radius $r$ (Einasto profiles)
tends to be favoured by recent high-resolution N-body simulations
\citep{navarro04,2006AJ....132.2685M,2008MNRAS.391.1685S,2012MNRAS.tmp.2773M} and also by galaxy
observations \citep{2011AJ....142..109C}. Simulations including baryons and
feedback processes are important to address further the question of the (dark) matter
profile in the innermost region \citep[e.g.,][]{2012MNRAS.tmp.2773M}.}. We note that Einasto profiles
give slightly more `signal' than NFW halos, making our conclusions on
detectability conservative.

Cold DM N-body simulations show a high level of clumpiness in the DM distribution
\citep[e.g.,][]{2007ApJ...657..262D,2008MNRAS.391.1685S}. These substructures boost
the signal in the outer parts of the DM halos. In agreement with the analysis of
\citet{2012MNRAS.419.1721G}, we find that the boost in galaxy clusters is larger
than the boost obtained for less massive objects such as dSphs. For the latter,
boost are $\lesssim 2$ \citep{2011MNRAS.418.1526C}, whereas we obtain an overall
boost of $\sim 10-20$ for galaxy clusters  based on conservative assumptions
for the substructure parameters (the impact of these parameters in discussed in
Section~\ref{sec:impact_subpars}). The reason is twofold: first, dSphs are
less massive so that the mass range of substructures is smaller (the minimal mass
is assumed to be the same regardless of the object), hence the number of objects,
and their overall contribution; second, the effective angular size on the sky is
larger for dSphs so that current instruments integrating out to $0.5^\circ$
integrate less substructure signal (see also \citealt{2012MNRAS.419.1721G}). 
These boost are obtained from the following configuration|used throughout the paper
 with the exception of Section~\ref{sec:impact_subpars}|for
the mass and spatial distribution of the substructures : i) $dN_{\rm
subs}/dM\propto M^{-1.9}$ with a mass fraction $f=10\%$ in substructures
\citep{2008MNRAS.391.1685S}, a minimal and maximal mass of $10^{-6}~M_\odot$ and
$10^{-2}M_{\rm cluster}$ respectively, and the \citet{2001MNRAS.321..559B}
concentration (down to the minimal mass); ii) the substructure spatial distribution
$dN_{\rm subs}/dV$  follows the host halo smooth profile.  For this configuration,
we checked
that the boost 
is only mildly dependent on this mass by varying the mass from $10^{-6}$ to 1~$M_\odot$.
We note that the minimal mass for sub-halos can be as small as
$10^{-10}~M_\odot$ depending on the particle physics model (see
\citealt{2006PhRvL..97c1301P}, and references therein). 

A complete study of the boost should consider different profiles, different
parametrisations for the mass-concentration relationship, etc. This will
be fully addressed in a future work.  However, given the impact it can have
on the $\langle\sigma_{\rm ann}v\rangle$ limit (or detectability for current
instruments), a short discussion and general trends are given in Section~\ref{sec:impact_subpars}.

\section{$J$ factors for the MCXC sample}
\label{sec:clusterhalosample}

The MCXC \citep{piffaretti11} contains 1743 clusters of galaxies detected in
X-rays, and assembled from publicly available catalogues mainly based on the ROSAT
All Sky Survey or ROSAT serendipitous catalogues. Most observational constraints
and predictions are expressed in terms of $\Delta=500$ or $\Delta=200$. For
instance, the mass of a halo, $M_\Delta$ can be defined within a radius $R_\Delta$
within which the average density reached $\Delta$ times the critical density of the
Universe (at a given redshift). The MCXC provides homogenised quantities for each
clusters computed within $\Delta=500$, e.g., the standardised [0.1-2.4]~keV X-ray
luminosity $L_{500}$, the total mass $M_{500}$, the radius $R_{500}$. 

To fully describe the NFW profile parameters (see Eq.~\ref{eq:rhoNFW}) for each
galaxy cluster of the MCXC catalogue, we used the provided $M_{500}$ together with a
mass-concentration relationship (i.e., $c_\Delta$ is fully determined by the cluster
mass $M_\Delta$). This relation is observationally constrained at the cluster scale
\citep{2005A&A...435....1P,2007ApJ...664..123B,2010A&A...524A..68E}. It has also
been shown to depend on the epoch of halo formation by numerical simulations of
structure formation
\citep{2001MNRAS.321..559B,dolag04,2008MNRAS.390L..64D,2011ApJ...740..102K}.
Although the data present a large dispersion, a systematic offset remains
unexplained \citep{2008MNRAS.390L..64D,2010MNRAS.405.2161D}. In this study, we
assume the \citet{2008MNRAS.390L..64D} mass-concentration relation. 

For an NFW profile $r_s=R_\Delta/c_\Delta$ and the scale density $\rho_s$ is
obtained from the mass measurement. The $J$ factor for all clusters are then
calculated from Eqs.~(\ref{eq:J}) and (\ref{eq:rhoNFW}) with {\sc clumpy}.

\subsection{Brightest targets}
\label{sec:thecatalogue}

\begin{figure*}
\begin{center}
\includegraphics[width=0.49\linewidth]{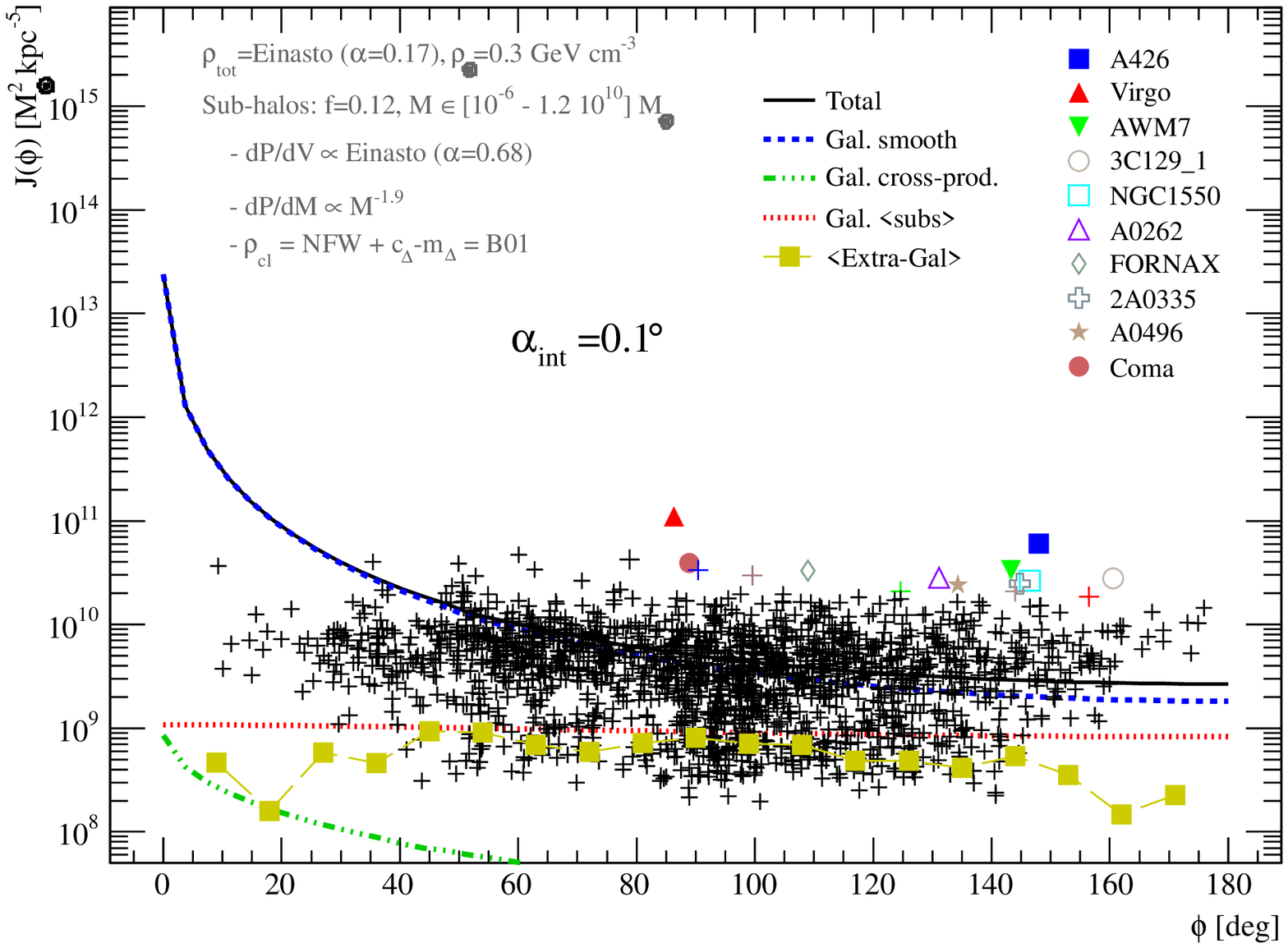}
\includegraphics[width=0.49\linewidth]{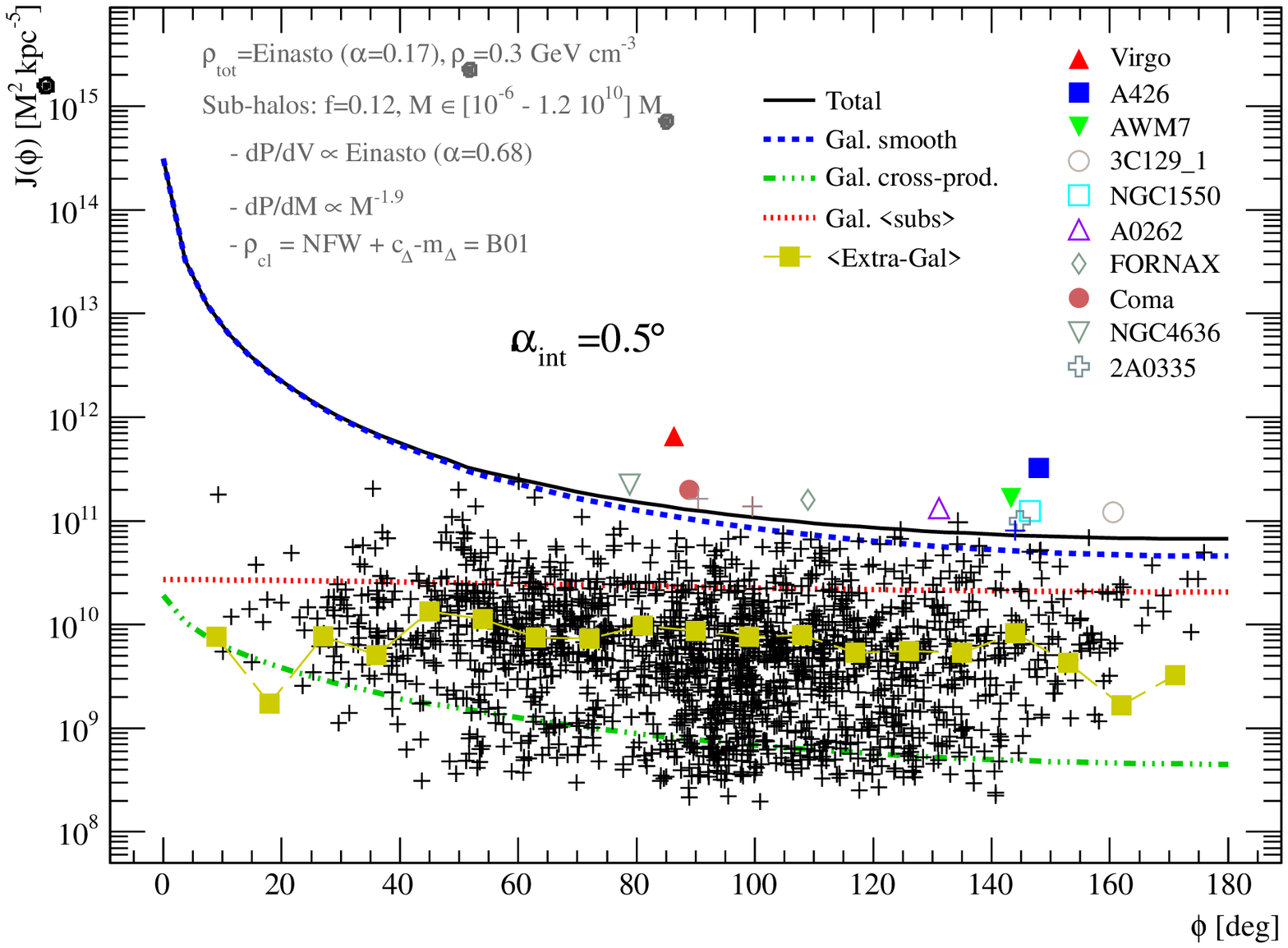}
\caption{Computed $J$-factors for the MCXC sources (the 10 highest-contrast
  clusters are highlighted, the remaining are shown with a `+' symbol) vs Galactic
  DM background (total is the sum of smooth, sub-halos, and cross-product|see
  details in \citealt{2012CoPhC.183..656C}). The yellow filled square symbols are
  evaluated from the cumulative of the cluster signal in different $\phi$ bins:
  this can be interpreted as a lower limit for the extra-galactic DM annihilation
  signal. {\bf Left panel:} integration angle $\alpha_{\rm int}=0.1^\circ$. {\bf
  Right panel:}  $\alpha_{\rm int}=0.5^\circ$.}
\label{fig:J_v_phi}
\end{center}
\vspace{-0.4cm}
\end{figure*}

\begin{table*}
\begin{center}
  \caption{Twenty brightest galaxy clusters from the MCXC and their
    contrast $J/J_{\rm Gal}$ for $\alpha_{\rm int}=0.1^\circ$. The DM
    Galactic background is evaluated at the position of the cluster
    (angle $\phi$ away from the Galactic centre, see
    Fig.~\ref{fig:J_v_phi}).}
\label{tab:tab1}
\begin{tabular}{@{}lccccccccccll} \hline\hline
  Name                  &Index  & $l$  &  $b$ & $\phi$&   $d$    & $\!\!\!\!\log_{10} \left(\frac{M_{\rm tot}}{1 M_\odot}\right)\!\!\!\!$ &   $\alpha_{80\%}$ &\multicolumn{3}{c}{$\!\!\log_{10}\left[J(\alpha_{\rm int})/(M_\odot^2~{\rm kpc}^{-5})\right]\!\!$} & \multicolumn{2}{c}{$\frac{J(\alpha_{\rm int})}{J_{\rm Gal}(\alpha_{\rm int})}$[rank]$^\ddagger$}\\
 &MCXC   &(deg) &(deg) & (deg) &     (Mpc)& $-$  &  (deg) &   $(0.1^\circ)$ &  $(0.5^\circ)$ & $\!\!(\alpha_{80\%})\!\!$ &   $(0.1^\circ)$ &  $(0.5^\circ)$ \vspace{0.0cm}\\
\hline
  Virgo                 &884    &283.8 & 74.4 & 86.3&  15.4  & 14.3  & 3.3 & 11.1 & 11.8  & 12.6 & \!\!\!20.7 [2]   & 4.9 [1]           \vspace{0.02cm}\\
  A426                  &258    &150.6 &-13.3 &148.0&  75    & 15.1  & 1.2 & 10.8 & 11.5  & 11.8 & \!\!\!21.2 [1]   & 4.5 [2]           \vspace{0.02cm}\\
 {\bf A3526}$^\star$    &915    &302.4 & 21.6 & 60.1&  48.1  & 14.5  & 1.2 & 10.7 & 11.4  & 11.7 & {\bf 4.7} [30]   & {\bf 0.9} [17]    \vspace{0.02cm}\\
 NGC 4636               &906    &297.7 & 65.5 & 78.9&  13.2  & 13.3  & 1.7 & 10.6 & 11.4  & 11.8 & 6.8 [13]         & 1.4 [9]           \vspace{0.02cm}\\
{\bf \em A3627}$^{\star,\diamond}$&1231&325.3&-7.1&35.4& 66  & 14.6  & 0.9 & 10.6 & 11.3  & 11.5 &{\bf \em 1.4} [-] & {\bf \em 0.3} [-] \vspace{0.02cm}\\
  Coma                  &943    & 57.2 & 88.0 & 88.9&  96.2  & 14.9  & 0.8 & 10.6 & 11.3  & 11.5 & 7.7 [10]         & 1.6 [8]           \vspace{0.02cm}\\
{\bf NGC5813}$^\star$   &1147   &359.2 & 49.8 & 49.8&  21.3  & 13.6  & 1.4 & 10.6 & 11.3  & 11.6 &{\bf 2.7} [-]     & {\bf 0.6} [39]    \vspace{0.02cm}\\
{\bf \em Ophiuchus}$^{\star,\diamond}$&1304&0.6&9.3&9.3&116. & 15.0  & 0.7 & 10.6 & 11.2  & 11.4 &{\bf \em 0.1} [-] & {\bf \em 0.02}[-] \vspace{0.02cm}\\
{\bf NGC5044}$^\star$      &978    &311.2 & 46.1 & 62.8&  36.9  & 14.0  & 1.0 & 10.5 & 11.2  & 11.5 &{\bf 3.6} [-]     & {\bf 0.7} [-]     \vspace{0.02cm}\\ 
    AWM7                &224    &146.3 &-15.6 &143.3&  72.1  & 14.5  & 0.8 & 10.5 & 11.2  & 11.3 & \!\!\!11.5 [3]   & 2.3 [3]           \vspace{0.02cm}\\ 
     A1060              &689    &269.6 & 26.5 & 90.4&  53.1  & 14.2  & 0.9 & 10.5 & 11.2  & 11.4 & 6.7 [14]         & 1.3 [11]          \vspace{0.02cm}\\
Fornax                  &285    &236.7 &-53.6 &109.0&  21.7  & 13.5  & 1.2 & 10.5 & 11.2  & 11.5 & 8.6 [7]          & 1.6 [7]           \vspace{0.02cm}\\
 A1367                  &792    &235.1 & 73.0 & 99.6&  89.3  & 14.6  & 0.7 & 10.5 & 11.1  & 11.2 & 6.9 [12]         & 1.3 [12]          \vspace{0.02cm}\\ 
{\bf \em J1324.7-5736}$^\star,\diamond$&990&307.4&5.0& 52.8&79.5&14.5& 0.7 & 10.5 & 11.1  & 11.3 &{\bf\em 2.3} [-]    & {\bf \em 0.4}[-]      \vspace{0.02cm}\\
 A0262                  &158    &136.6 &-25.1 &131.1&  68.4  & 14.3  & 0.7 & 10.5 & 11.1  & 11.3 & 9.0 [6]          & 1.7 [6]           \vspace{0.02cm}\\
{\em 3C129}$^{\diamond}$&350    &160.5 &  0.3 &160.5&  91.7  & 14.5  & 0.7 & 10.4 & 11.1  & 11.2 &\!\!\!{\em 10.2} [4]&{\em 1.8} [4]    \vspace{0.02cm}\\
{\bf A2199}$^\star$     &1249   & 62.9 & 43.7 & 70.8&  12.4  & 14.7  & 0.6 & 10.4 & 11.0  & 11.1 &{\bf 3.5} [-]     & {\bf 0.6} [38]    \vspace{0.02cm}\\
 NGC1550                &324    &191.0 &-31.8 &146.5&  55.2  & 14.1  & 0.8 & 10.4 & 11.1  & 11.2 & 9.2 [5]          & 1.7 [5]           \vspace{0.02cm}\\
{\bf A3571}$^\star$     & 1048  &316.3 & 28.6 & 50.6& 159.7  & 14.9  & 0.5 & 10.4 & 11.0  & 11.0 & {\bf 1.9} [-]    & {\bf 0.3} [-]     \vspace{0.02cm}\\
2A0335                  & 286   &176.3 &-35.5 &144.8& 142.5  & 14.8  & 0.5 & 10.4 & 11.0  & 11.0 & 8.6 [8]          & 1.4 [10]          \vspace{0.0cm}\\
\hline
\end{tabular}
\\
$^\ddagger$ Whenever the rank is larger than 50, we use [-].\\  
$^\star$ Weakly contrasted {\bf clusters} are probably not the best targets.\\  
$^\diamond$ {\em Clusters} close to the Galactic plane are not favoured targets.  
\end{center}
\vspace{-0.4cm}
\end{table*}

Figure~\ref{fig:J_v_phi} provides a synthetic view of the $J$-factor for each
galaxy cluster of the MCXC catalogue as a function of their angle $\phi$ away from
the Galactic centre.  The integration angle is taken to be $\alpha_{\rm
int}=0.1^\circ$ (left panel) and $\alpha_{\rm int}=0.5^\circ$ (right panel), the
typical range of value for the energy-dependent angular resolution of current
$\gamma$-ray instruments such as Fermi-LAT (in the high-energy range above
$\sim$10~GeV) and H.E.S.S. Table~\ref{tab:tab1} gathers results for the twenty
brightest clusters in the MCXC. From this table, we simply note that $J$-factors
are competitive with those obtained for dSphs \citep[e.g.,][]{2011ApJ...733L..46W},
confirming that galaxy clusters are valid targets for dark matter annihilation
searches (see also \citealt{2011JCAP...12..011S,2012MNRAS.419.1721G}). The
panels of Fig.~\ref{fig:fig2} show a skymap version of Fig.~\ref{fig:J_v_phi}. The
top left panel shows the $J$ factor induced by DM annihilation in the Galactic halo
cumulated with all MCXC objects. The top right panel shows the $J$ factor skymap for
all MCXC galaxy clusters only. The bottom panel locates the twenty most promising
targets labelled by distance, absolute $J$-factor value and {\em contrast} with
respect to the DM Galactic signal.

Several clusters including Virgo, Coma, Fornax, {\sc ngc}~5813 and Ophiuchus have
already been credited to be interesting sources in numerous studies given
their masses and distances
\citep{2006A&A...455...21C,2009PhRvD..80b3005J,2011JCAP...12..011S,2011PhRvD..84l3509P,2012MNRAS.419.1721G,2012arXiv1201.1003H}. 
Other objects, such as, 3C129 and AWM7 were only highlighted from the HIFLUGCS
catalogue analysis
\citep{2009PhRvD..80b3005J,2011PhRvD..84l3509P,2012JCAP...01..042H}. With ten times
more objects, the MCXC gives a more exhaustive list of potential targets including,
e.g., J0123.6+3315 and J1324.7-5736 (see Table~\ref{tab:tab1} and
Fig.~\ref{fig:fig2}).

Some differences exist with previous calculations (see
App.~\ref{app2}). These can be partly attributed to a different
prescription for the substructures. However, another important
difference comes from the fact that almost all previous studies are
based on the $M_{500}$ values obtained from the HIFLUGCS catalogue
\citep{2002ApJ...567..716R,2007A&A...466..805C}. In particular, some
of `brightest' objects found (e.g., Coma, Fornax, AMW7) have larger
masses than those provided in the MCXC catalogue. As discussed in the
App.~A of \citet{piffaretti11}, the MCXC relies on a more accurate model
for the gas distribution, and many comparisons to numerical simulations
indicate that any systematic uncertainties are now $\lesssim15-20$\%
\citep{2008A&A...491...71P}.

\subsection{Galactic and extra-galactic DM background}
\label{sec:distribution}

\begin{figure*}
\begin{center}
\includegraphics[clip=,width=0.49\linewidth]{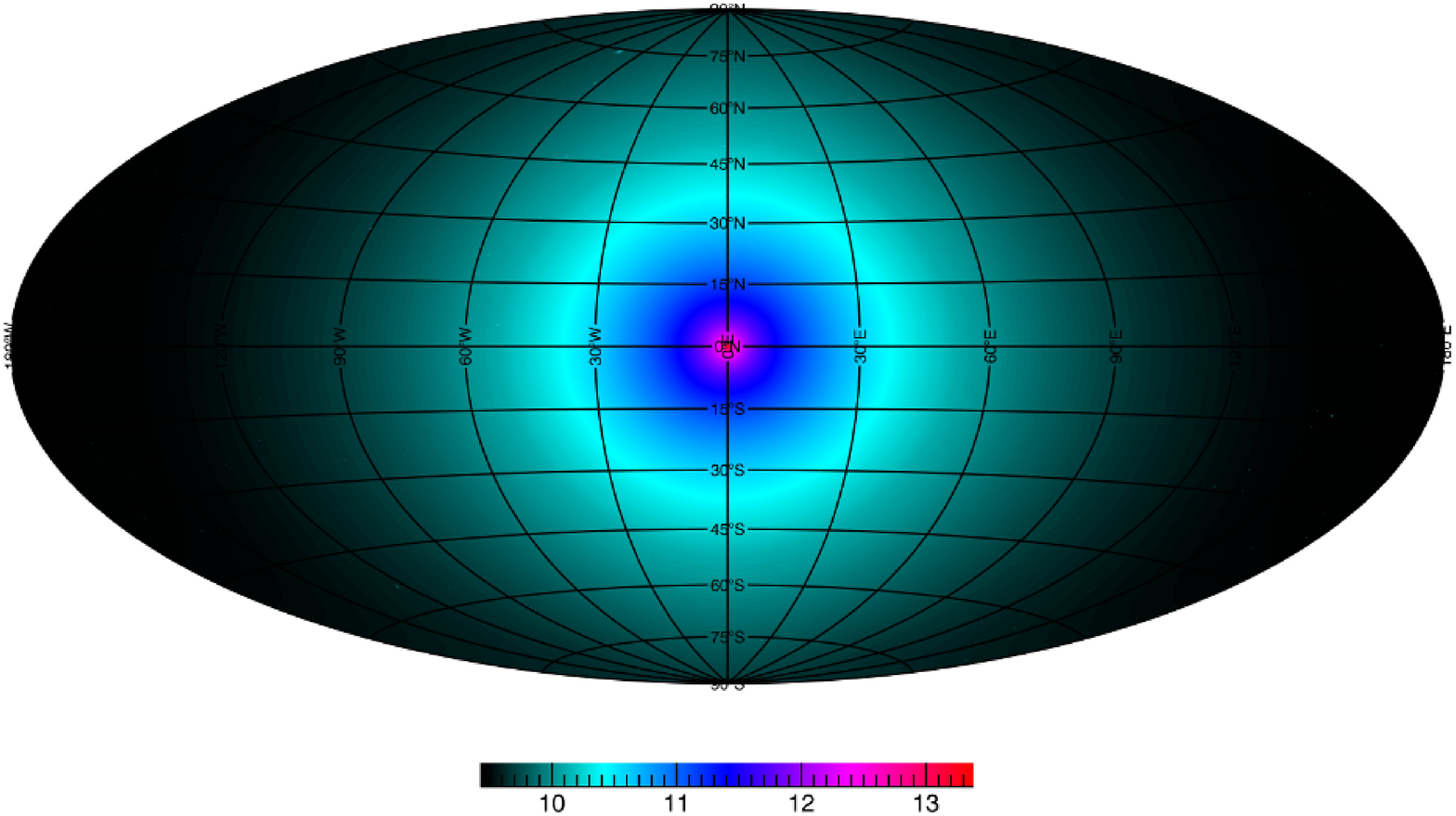}
\includegraphics[clip=,width=0.49\linewidth]{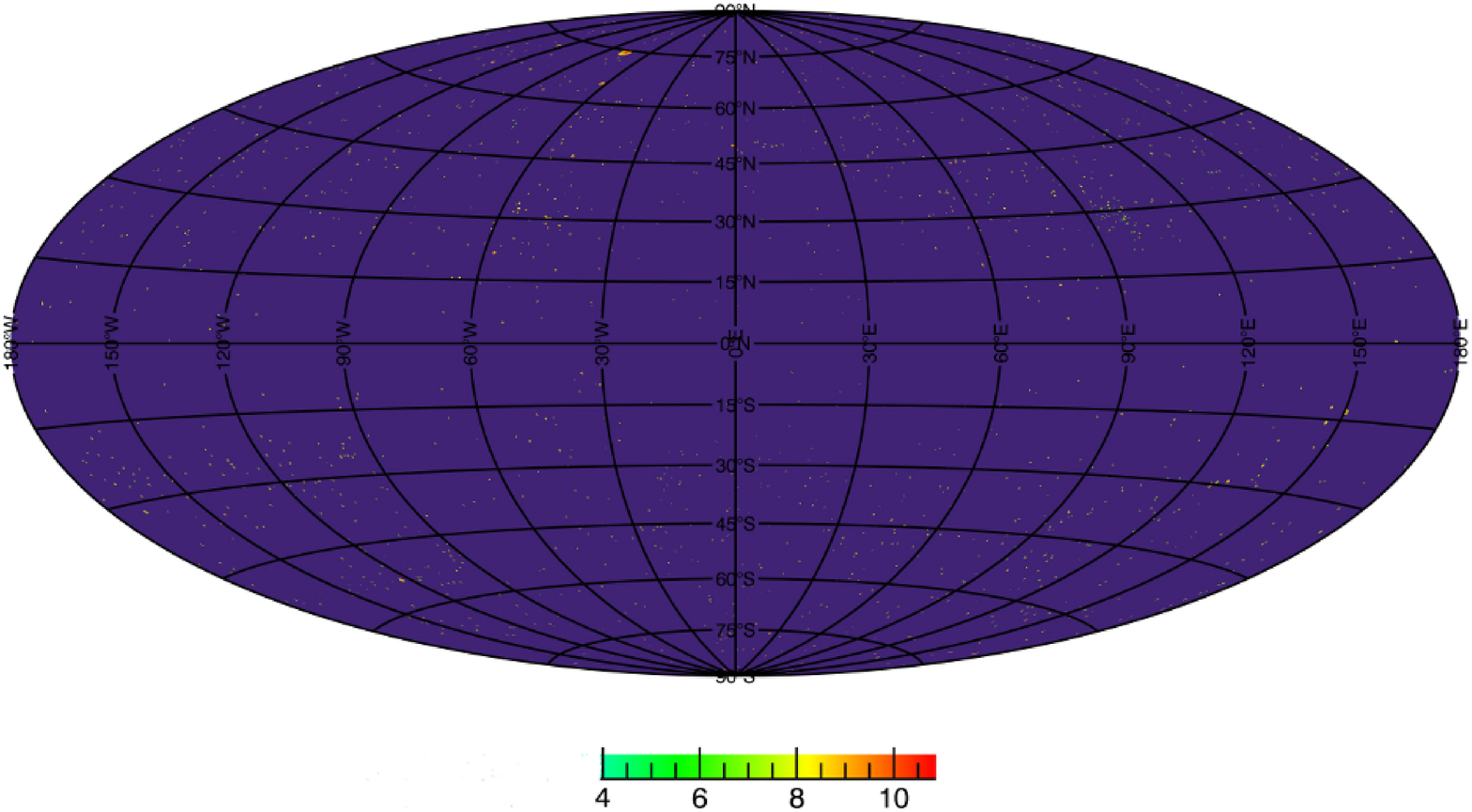}
\includegraphics[clip=,width=0.8\linewidth]{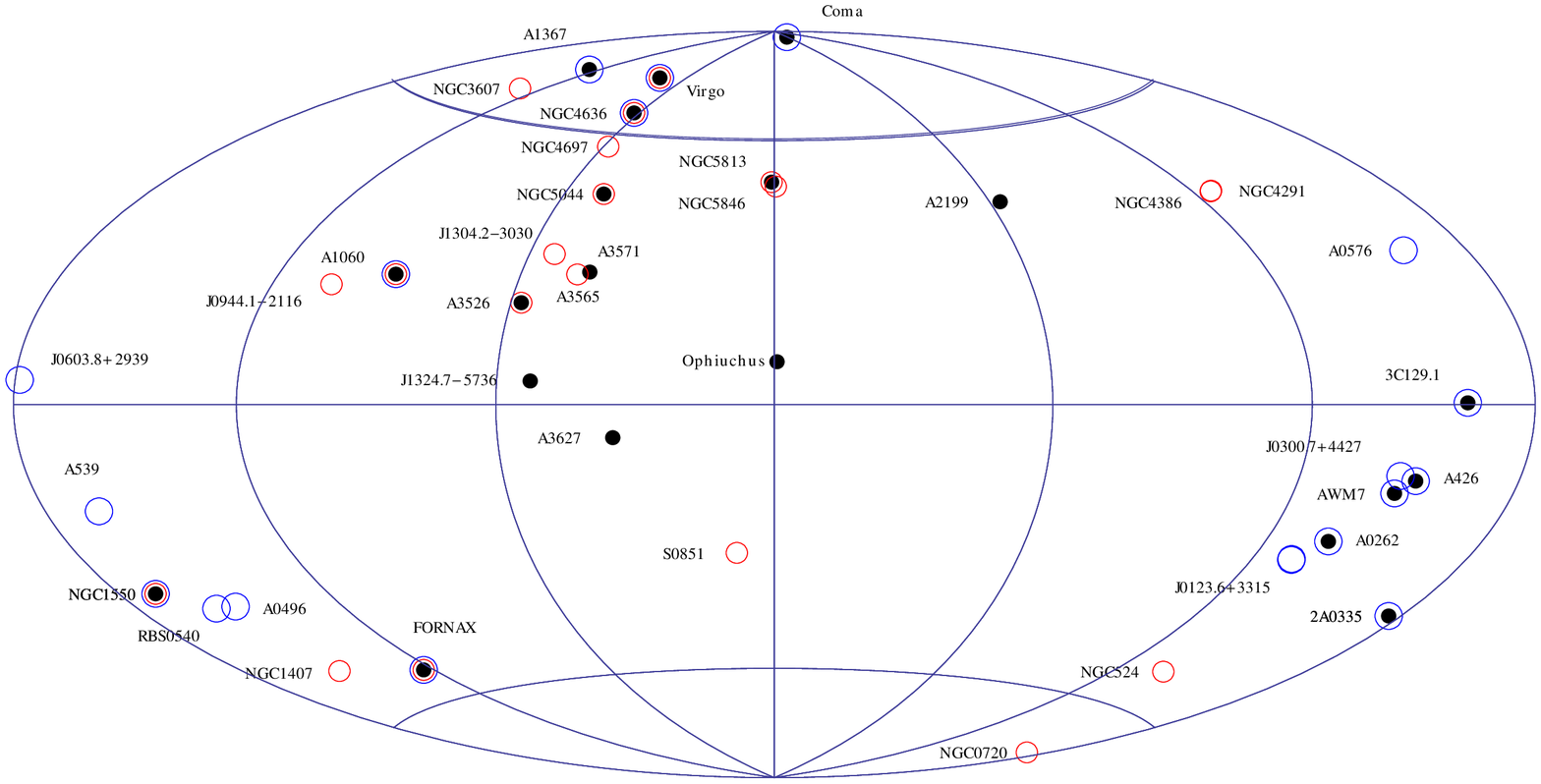}
\caption{{\bf Top panels:} $J$-factor skymap for $\alpha_{\rm int}=0.1^\circ$
 for Galactic + MCXC sources (left) and MCXC sources only (right). {\bf Bottom
 panel:} positions of the 20 closest (red circles), brightest (black points),
 and highest $J/J_{\rm Gal}$ (blue circles) from the MCXC.}
\label{fig:fig2}
\end{center}
\end{figure*}

Galactic DM provides a `diffuse' DM emission $J_{\rm Gal}$ that can drown the point-like
emissions we are looking for.  The value of the local DM density is still loosely
constrained in $[0.2 - 0.4]$~GeV~cm$^{-3}$ by several techniques
\citep{2009PASJ...61..227S,2010JCAP...08..004C,2010A&A...523A..83S,2011MNRAS.416.2318G,2011JCAP...11..029I}.
We assume here $\rho_\odot=0.3$~GeV~cm$^{-3}$. The value for $J_{\rm
Gal}(\phi\gtrsim 20^\circ)$ is also very sensitive to the Galactic sub-halo
distribution. The Galactic signal is thus uncertain by a factor of a few.  We
calculate in Table~\ref{tab:tab1} the {\em contrast}, i.e., the ratio between the
cluster signal to the DM Galactic signal. As shown in Figs.~\ref{fig:J_v_phi} and
\ref{fig:fig2}, the DM Galactic signal has a shallow latitudinal dependence  except
towards the Galactic centre ($\theta\gtrsim 5^\circ$) where the signal is maximal.
Several of the brightest sources are close to the galactic centre, namely
Ophiuchus, A3627(Norma), and J1324.7-5736. Although they exhibit a large
$J$-factor, their {\em contrast} is low, and they are not favoured. Indeed, the
contrast indicates when a point-like observation strategy becomes less promising
than a strategy based, e.g., on the detection of a gradient for smooth Galactic
halo towards the Galactic centre \citep[as done in][]{2011PhRvL.106p1301A}. Away
from the Galactic centre, we have $J_{\rm Gal}\propto\alpha_{\rm int}^2$.  This is
illustrated by the left and right panels of Fig.~\ref{fig:J_v_phi}, where the value
of $J_{\rm Gal}$ is multiplied by 25 moving from $\alpha_{\rm int}=0.1^\circ$ to
$\alpha_{\rm int}=0.5^\circ$. However, the corresponding signal from each cluster
is only marginally increased, meaning that the contrast is worsened for large
integration angles. 

The diffuse extra-Galactic DM signal constitutes another background,
the level of which has been estimated from N-body simulations (see,
e.g., Fig.~4 of \citealt{2011PhRvD..83b3518P}). It is not considered
here. However, by averaging in each $\phi$ bin the signal from all
clusters and correcting for the solid angle element, we derive a first
`data-driven' estimate of this extra-galactic contribution, and we
find $J_{\rm extra-gal}\gtrsim J_{\rm Gal}/5$ (yellow filled squares
in Fig.~\ref{fig:J_v_phi}). Larger samples of galaxy clusters are
required to refine this figure.

The five brightest sources in Table~\ref{tab:tab1} are located far from the
Galactic centre and plane, and therefore have the `best' {\em contrast} w.r.t.
diffuse DM and astrophysical emissions (located mostly in the
disk). These sources are also amongst the closest targets and have the largest
angular size. As we will show in the next sections, this will prove crucial
for the detection prospects once the astrophysical background and the angular
response of the instruments are taken into account.

\subsection{Distribution of $J$ factors and $\alpha_{80\%}$ for the cluster sample}

Most of the galaxy clusters in the MCXC are faint objects (see
Fig.~\ref{fig:J_v_phi}). A stacking analysis is appealing if the slope of $\log N
- \log J$ is steeper than $-1$, indicating that the number of sources increases
more rapidly than the brightness of those sources diminishes. 

\begin{figure*}
\begin{center}
\includegraphics[width=0.35\linewidth]{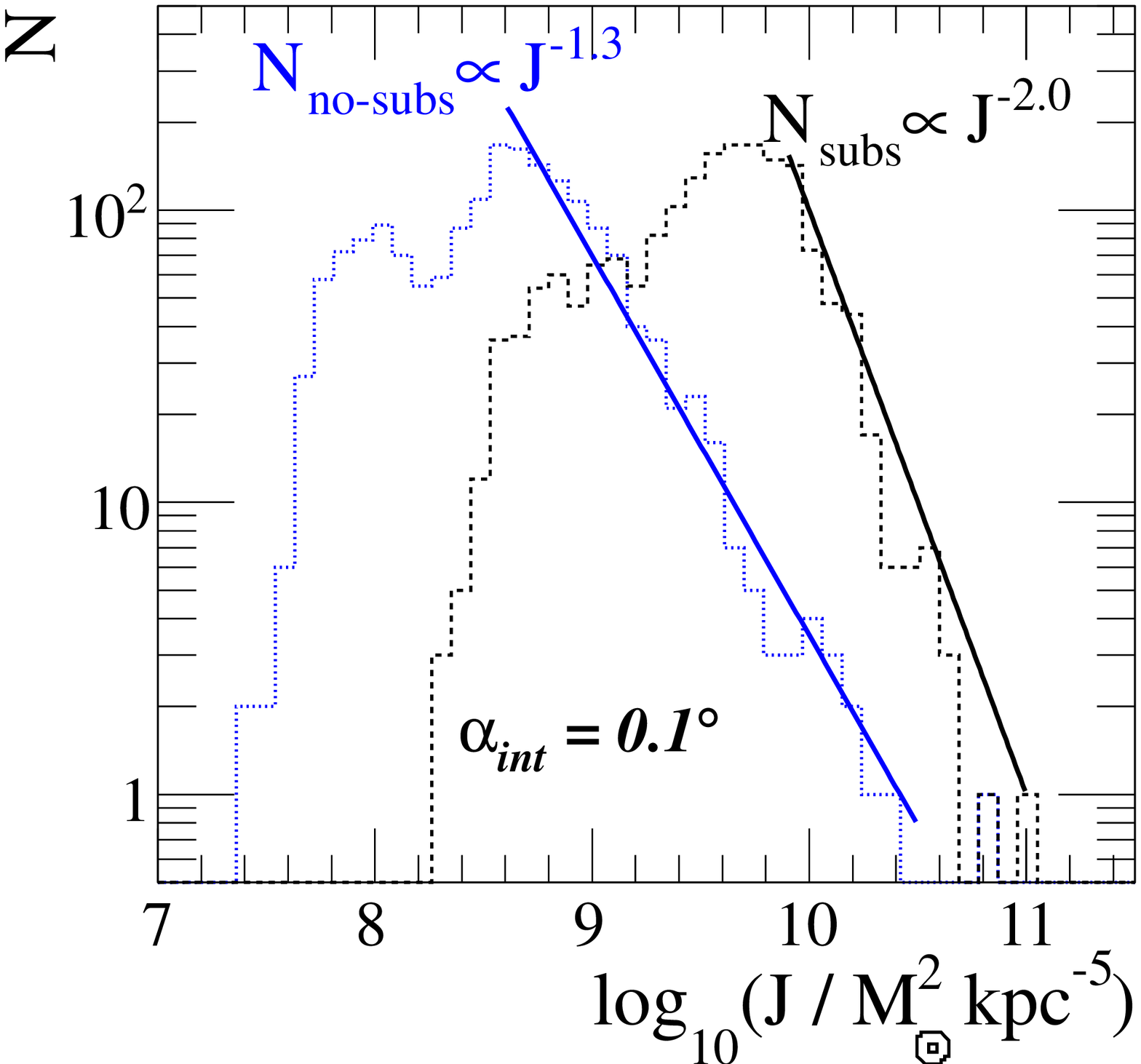}\hspace{2cm}
\includegraphics[width=0.35\linewidth]{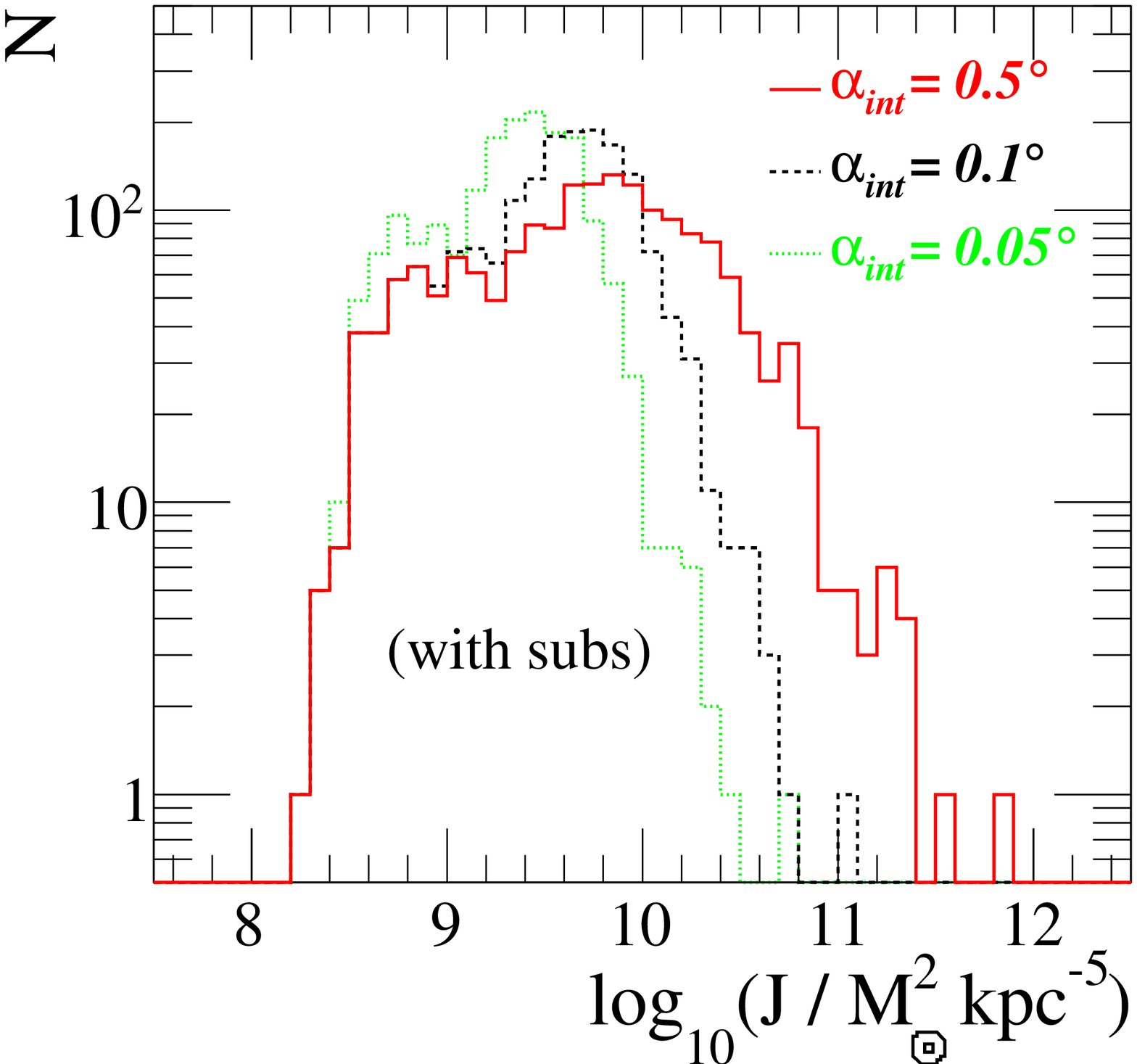}
\includegraphics[width=0.35\linewidth]{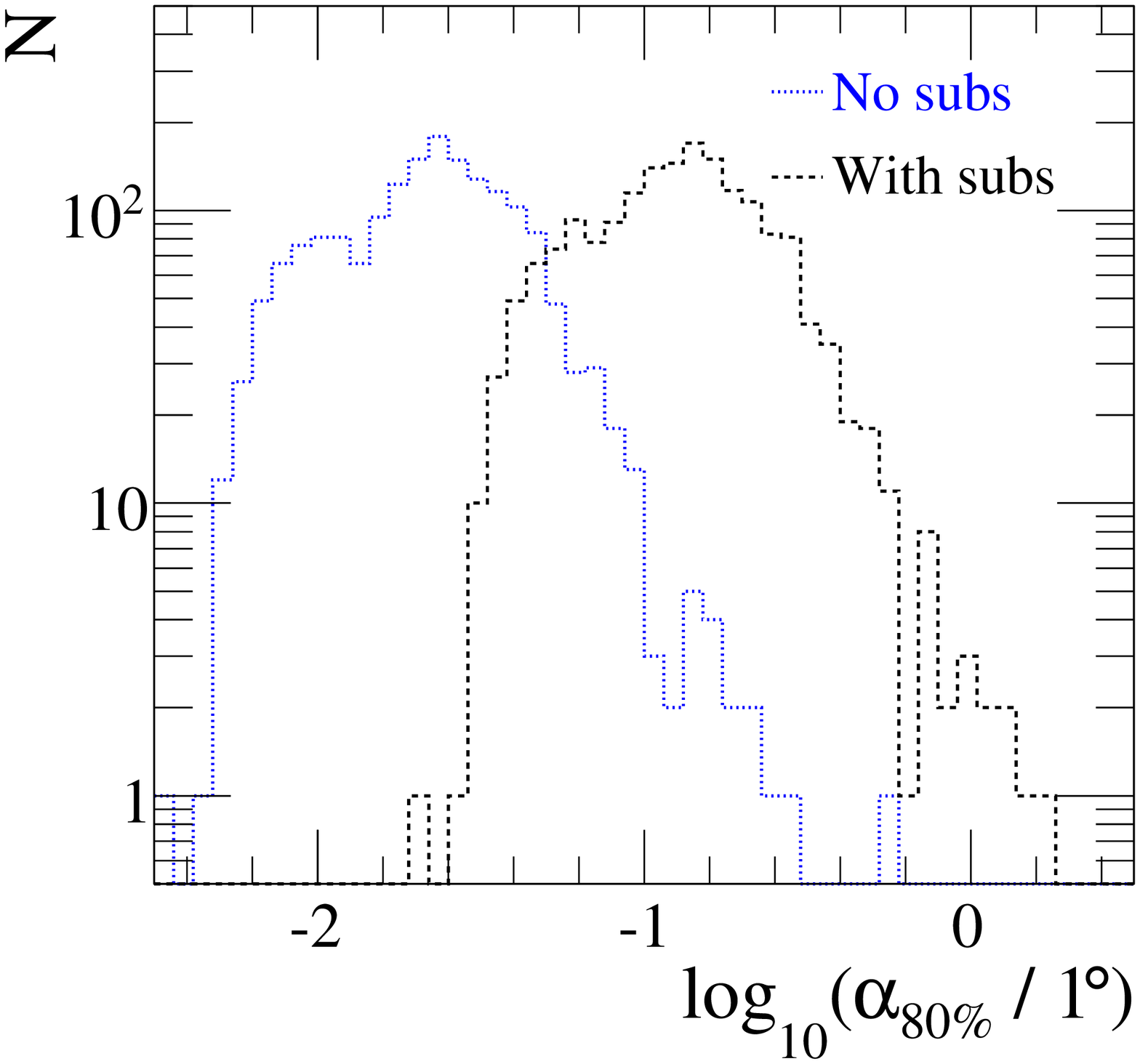}\hspace{2cm}
\includegraphics[width=0.35\linewidth]{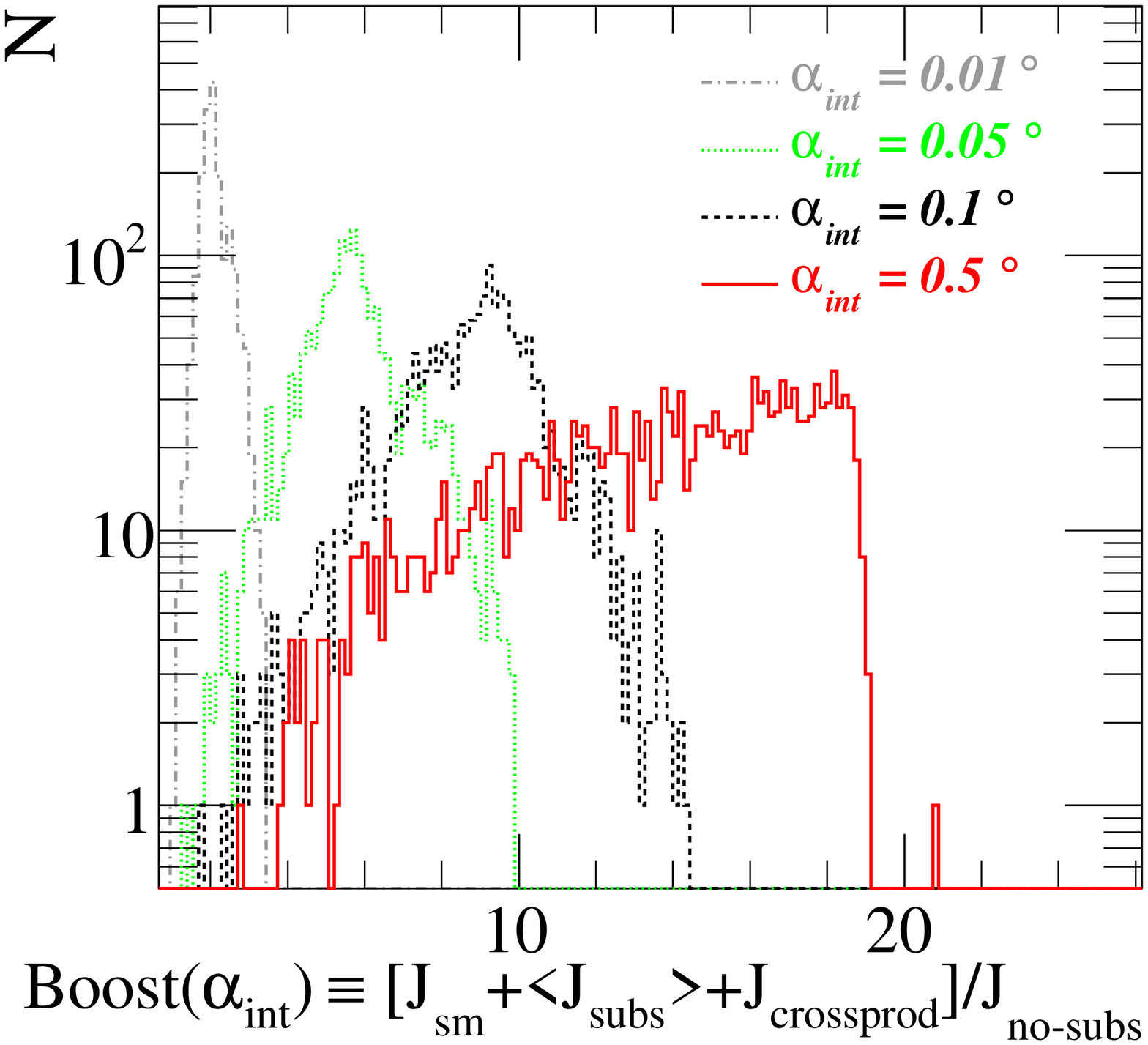}
\caption{{\bf Top panels:} the distribution of $\log_{10} N - \log_{10} J(\alpha_{\rm int})$.
The left panel shows a comparison without (dotted blue line) and with (dashed black
line) substructures for  $\alpha_{\rm int}=0.1^\circ$. The right panel shows  $\log_{10}
N - \log_{10} J(\alpha_{\rm int})$ for three different integration angles (all with
substructures). The solid lines are power-law fits on the brightest $J$ values of
the histograms.  {\bf Bottom left panel:} the distribution of $\alpha_{80\%}$ (the
integration angle containing 80\% of $J$) without (dotted blue line) and with
(dashed black line) substructures.  {\bf Bottom right panel:} the distribution of
boost factors (for the MCXC sample) for four integration angles. The boost is
defined to be the ratio of the total $J$ factor (with substructures) to the $J$
factor obtained in the hypothetical case where no substructures (only
smooth) exist in the galaxy cluster. }
\label{fig:fig3}
\end{center}
\end{figure*}

The $\log_{10} (N) - \log_{10} (J)$ distribution is shown in the top panels of
Fig.~\ref{fig:fig3}. We note that the double-peaked structure found is an
indication that the MCXC is neither complete nor uniform at high redshift. The top
left panel emphasises the importance of substructures for
$\alpha_{\rm int}=0.1^\circ$: in their absence (dotted blue line), we have $N_{\rm
no-subs}\propto J^{-1.3}$ such that there are $\gtrsim 20$ times more objects each
time $J$ is decreased by a factor of ten. With substructures (dashed red line) the
prospects for stacking are improved; $N_{\rm subs}\propto J^{-2.0}$ such that there
is now a factor $100$ increase in the number of target objects for the same factor
ten $J$ decrease. The lower left panel of Fig.~\ref{fig:fig3} shows the $\log_{10}
(N) - \log_{10} (\alpha_{\rm 80\%})$ distribution, where $\alpha_{80\%}$ is the
integration angle for which 80\% of the total $J$-factor is included. The quantity
$\langle\alpha_{80\%}\rangle$ is of importance as it corresponds to the desired
PSF in order to include most of $J$ in the majority of sources. This plot again
emphasises the role of substructures. The mean for the $\alpha_{80\%}$ distribution
moves from $\sim 0.03^\circ$ (dotted blue line) to $\sim 0.15^\circ$ when the
contribution of substructures is taken into account. This is more favourable
for current observatories, the angular resolution of which being at best
$\sim 0.1^\circ$.

The choice for the integration angle also impacts on the $\log_{10} (N) - \log_{10}
(J)$ distribution. The top right panel shows that larger integration angles have an
impact only for halos not fully encompassed, i.e., for the closest/brightest ones.
Indeed, objects whose $\alpha_{80\%}<\alpha_{\rm int}$ do not have significantly more signal when
$\alpha_{\rm int}$ is increased. For the bigger objects the interplay between the
different angular dependence of the smooth and substructure contributions shapes the
$\log_{10} (N) - \log_{10} (J)$ distribution.  The distribution of boost (for
different integration angles) shown in the bottom right panel of
Fig.~\ref{fig:fig3} illustrates this interplay. For very small $\alpha_{\rm int}$
(e.g., $0.01^\circ$ dash-dotted grey line), the signal from the smooth dominates
and the distribution is strongly peaked around 1 (no boost). As the integration
angle is increased ($0.5^\circ$, solid red line), the distribution is broadened,
asymmetric, and reaches a maximum of $\sim 20$.

\subsection{Impact of varying substructure parameters}
\label{sec:impact_subpars}

 As already underlined, several ingredients for the DM distributions (smooth and
subhalos) can affect the results above. For instance, physical processes involving
baryonic matter (such as AGN feedback) may produce a core distribution
\citep{2012MNRAS.tmp.2773M}. This could decrease the total $J$ values. However, for
galaxy clusters, a dominant part of the signal comes from substructures for
$\alpha_{\rm int}\gtrsim 0.05^\circ$ (as seen in bottom-right Fig.~\ref{fig:fig3}),
the distribution of which impacts the results significantly.
First, the smallest protohalo mass remains unknown, and it strongly
depends on the details of the dark matter candidate microphysics at the kinetic
decoupling
\citep[e.g.,][]{2005JCAP...08..003G,2006PhRvL..97c1301P,2009NJPh...11j5027B,2012arXiv1205.1914G}.
Second, the subhalo spatial distribution is found to be less concentrated than the
smooth halo one, and consistent with Einasto profiles in the recent Aquarius
\citep{2008MNRAS.391.1685S} and Phoenix \citep{2012arXiv1201.1940G} simulations. In
the latter the mass distribution slope $\alpha_{\rm M}$ ($dP/dM\propto M^{-\alpha_{\rm M}}$) is also
found to be steeper and close to 2, leading to a larger fraction of substructures 
in clusters than in galaxies.  When the slope is close to 2, the
contribution to the signal of small subhalos becomes as important as that of
larger ones, which can strongly boost the overall signal depending on the chosen
$c_\Delta-M_\Delta$ (concentration-mass) relation. 

Many studies have focused on the caracterisation (e.g., mean and variance, environment
effects) of this relation, but are limited by the mass resolution of currently
available numerical simulations. The state-of-the-art studies on galaxy clusters
apply down to a minimum halo mass $\sim 10^{10} M_\odot$
\citep{2002ApJ...568...52W,2006ApJ...652...71W,
2003ApJ...597L...9Z,2009ApJ...707..354Z,2007MNRAS.381.1450N,2011MNRAS.410.2309G,
2008MNRAS.387..536G,2008MNRAS.391.1940M,2010MNRAS.404..502G,2011MNRAS.411..584M,
2011ApJ...740..102K}. Extrapolations to the smallest subhalo mass are provided in a very
few studies only. For instance, almost all analyses of the DM annihilation signal are
based on two different parameterisations
\citep{2001MNRAS.321..559B,2001ApJ...554..114E}. \citet{2012MNRAS.422..185G} 
recently provided a new parameterisation of $c_\Delta-M_\Delta$: at $z=0$,
it is consistent with the \citet{2001MNRAS.321..559B} parametrisation in the  $\lesssim
10^{10} M_\odot$ range, but its redshift dependence is different.

\begin{figure}
\begin{center}
\includegraphics[width=0.9\linewidth]{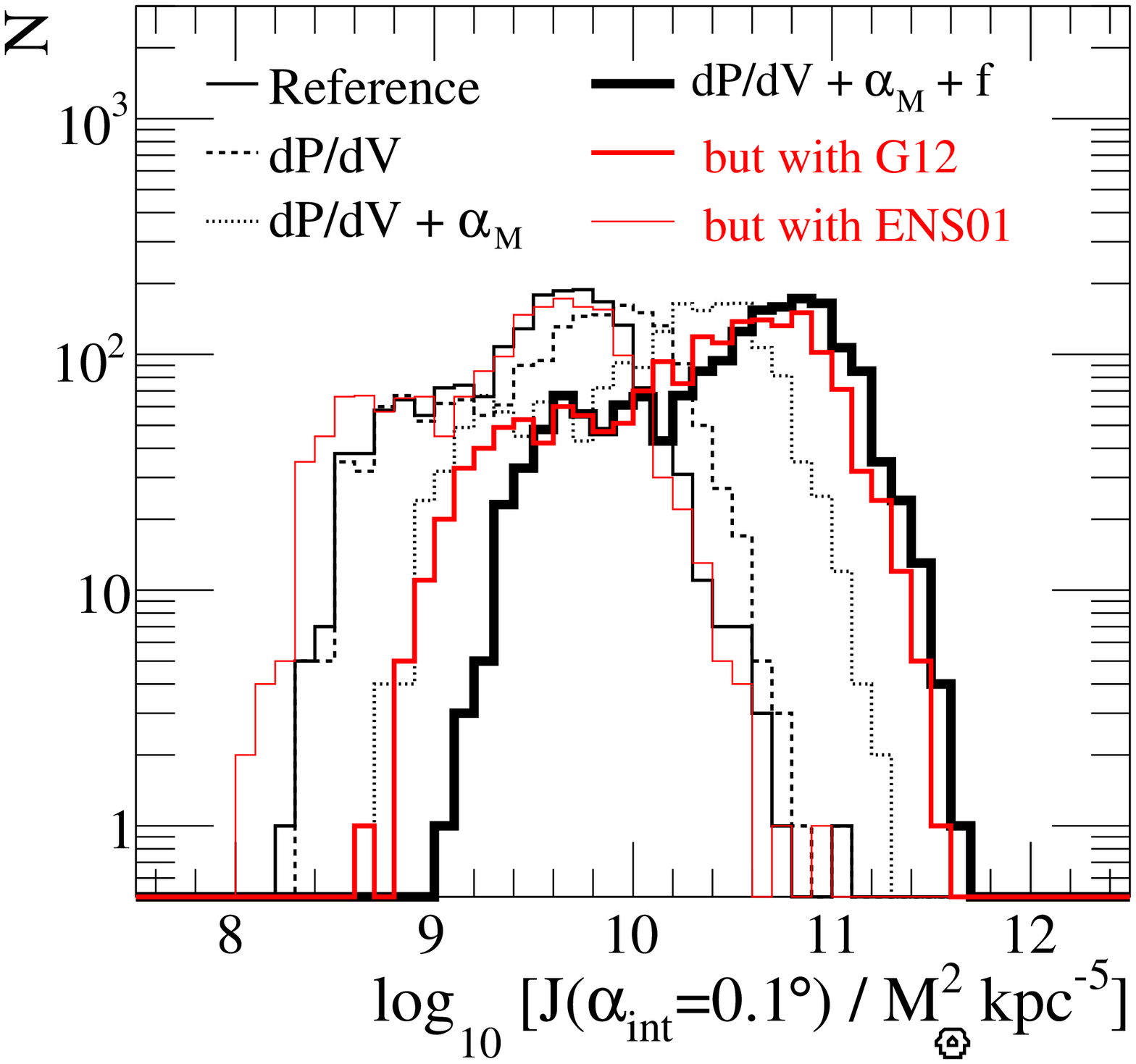}
\includegraphics[width=0.9\linewidth]{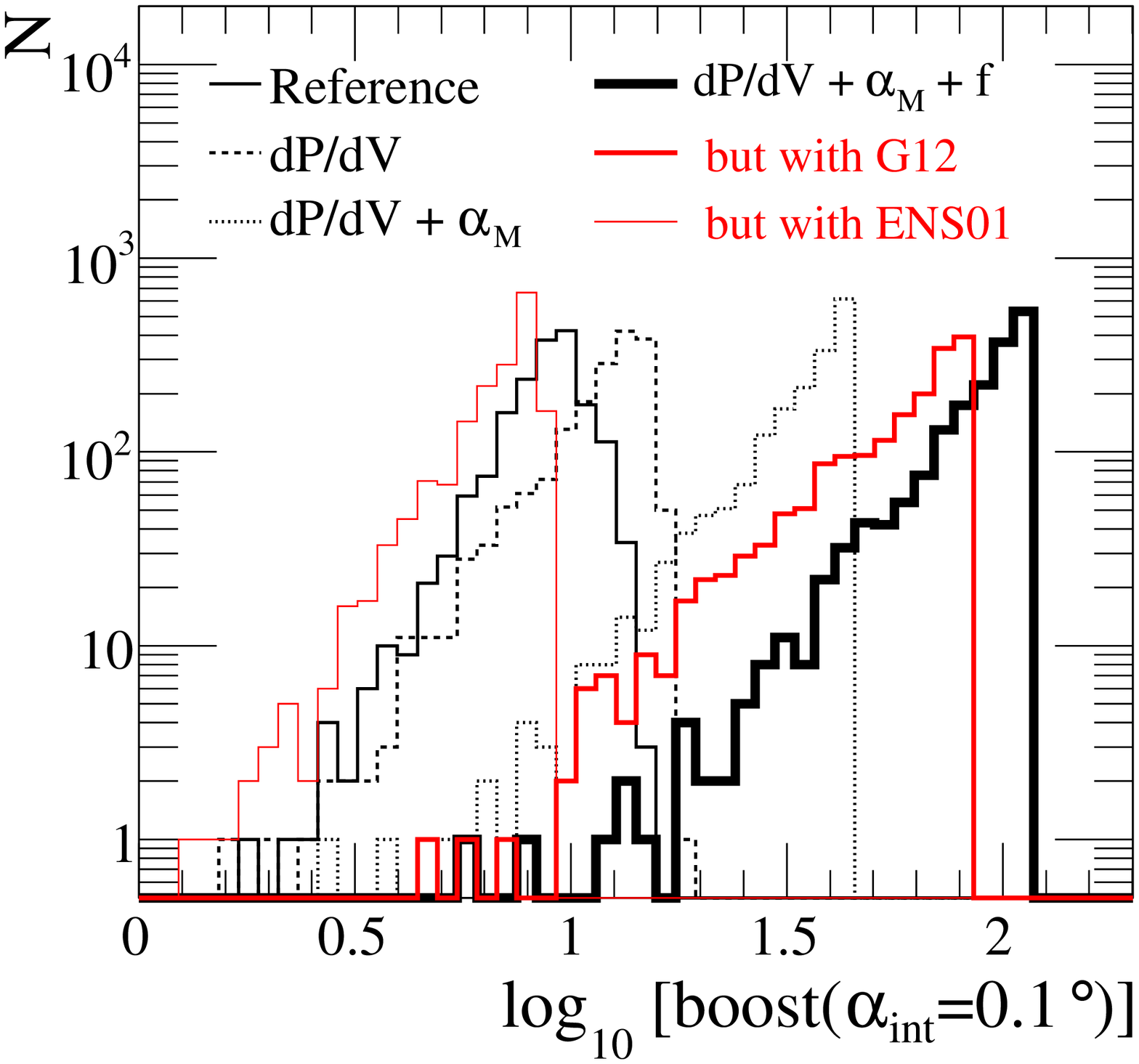}
\caption{The distribution of $\log_{10} N - \log_{10} J(\alpha_{\rm int})$ (top panel)
and boost factors (bottom panel) for different substructure configurations
(see text for details).}
\label{fig:fig3bis}
\end{center}
\end{figure}

Given the variety of results found in the literature and the uncertainties on some
parameters, we only select a few configurations below.
  \begin{itemize}
     \item $d{\cal P}/d{\cal V}_{\rm Phoenix}$: uses spatial distribution and scale
     radius as provided by the Phoenix project \citep{2012arXiv1201.1940G}, instead of
     following that of the smooth profile (all other parameters as in
     Section~\ref{sec:ingred_DM});
     \item $d{\cal P}/d{\cal V}_{\rm Phoenix}$ and $\alpha_{\rm M}=1.98$: uses a
     steeper slope for the mass distribution as found in Phoenix
     \cite{2012arXiv1201.1940G} instead of 1.9 (note that $\alpha_{\rm M}=1.94$ in
     \citealt{2008MNRAS.391.1685S});
     \item $d{\cal P}/d{\cal V}_{\rm Phoenix}$, $\alpha_{\rm M}=1.98$, and $f=0.3$:
     uses a DM mass fraction as found in Phoenix \citep{2012arXiv1201.1940G} instead of
     0.1 (as found in \citealt{2008MNRAS.391.1685S}).
     \item $d{\cal P}/d{\cal V}_{\rm Phoenix}$, $\alpha_{\rm M}=1.98$, $f=0.3$,
     and $(c_\Delta-M_\Delta)_{\rm ENS01}$ or $(c_\Delta-M_\Delta)_{\rm G12}$:
     uses a mass-concentration relation from \citet{2001ApJ...554..114E} and
     \citet{2012MNRAS.422..185G}, instead of using \citet{2001MNRAS.321..559B}.
  \end{itemize}
The impact is shown in Fig.~\ref{fig:fig3bis} for the $\log_{10} (N) - \log_{10}
(J)$ (top panel) and the $\log_{10} (N) - \log_{10}
({\rm Boost})$ (bottom panel) distributions for $\alpha_{\rm int}=0.1^\circ$.
Taking an Einasto profile for the spatial distribution of subhalos has only a minor
effect (solid thin- vs dashed-black line). A major impact is that of the value of
the parameter $\alpha_{\rm M}$ (solid thin- vs dotted-black line). It increases the
boost (and thus the signal) by about one order of magnitude. This increase can be
larger if a smaller minimal mass for the subhalos is chosen, but it can also be
decreased by a factor of ten if the minimal mass allowed is $10^3 M_\odot$.
Although the Phoenix and Aquarius simulations tend to prefer values close to 2, the
result is intrinsically limited by the mass resolution of the simulation, and this
slope can still be smaller at lower masses. Moreover, \citet{2009MNRAS.395.1950E}
argue that this slope could be overestimated, even in the simulation mass range.
Another obvious effect is from the mass fraction $f$ in substructures (thin vs
thick solid black line). Finally, the two red curves show the impact of the
$c_\Delta-M_\Delta$ parametrisation (thin and thick red lines) compared to using
\citet{2001MNRAS.321..559B} parametrisation (thick solid black line). Choosing
\citet{2012MNRAS.422..185G} gives slightly less signal, whereas using
\citet{2001ApJ...554..114E} washes out the boost completely.
To conclude, we see that the main source of uncertainties correspond to the slope
of the mass function, the minimal mass of the subhalos, and the concentration.
Unfortunately, these are also the least-constrained parameters from the available
numerical simulations. 

Our reference configuration gives a conservative estimate of the signal expected
from galaxy clusters. We underline that the shape of $\log_{10} N - \log_{10}
J(\alpha_{\rm int})$ is only weakly impacted by choosing other configuration (and
the ordering of the best targets|not shown|also remains mostly unaffected).
Therefore, the conclusions that will be reached below for the  stacking analysis
using the reference configuration hold regardless of this choice.  However, the
consequences of a larger boost would be the following: i) a larger signal and  a
better $\langle\sigma v\rangle$ limit set from non-detection, ii) an increase of
the 80\%-containing angle, iii) an enhanced contrast with respect to the Galactic
background resulting in an increased extragalactic to Galactic signal ratio.

\section{Halo stacking and results}
\label{sec:stacking}

There are two primary considerations for the MCXC stacking analysis: how to order
the sources, and how many sources to stack. This is discussed for different
situations before moving to the detection prospects for the stacking strategy.

\subsection{Strategy for a `perfect' instrument}

\begin{figure}
\begin{center}
\includegraphics[width=\linewidth]{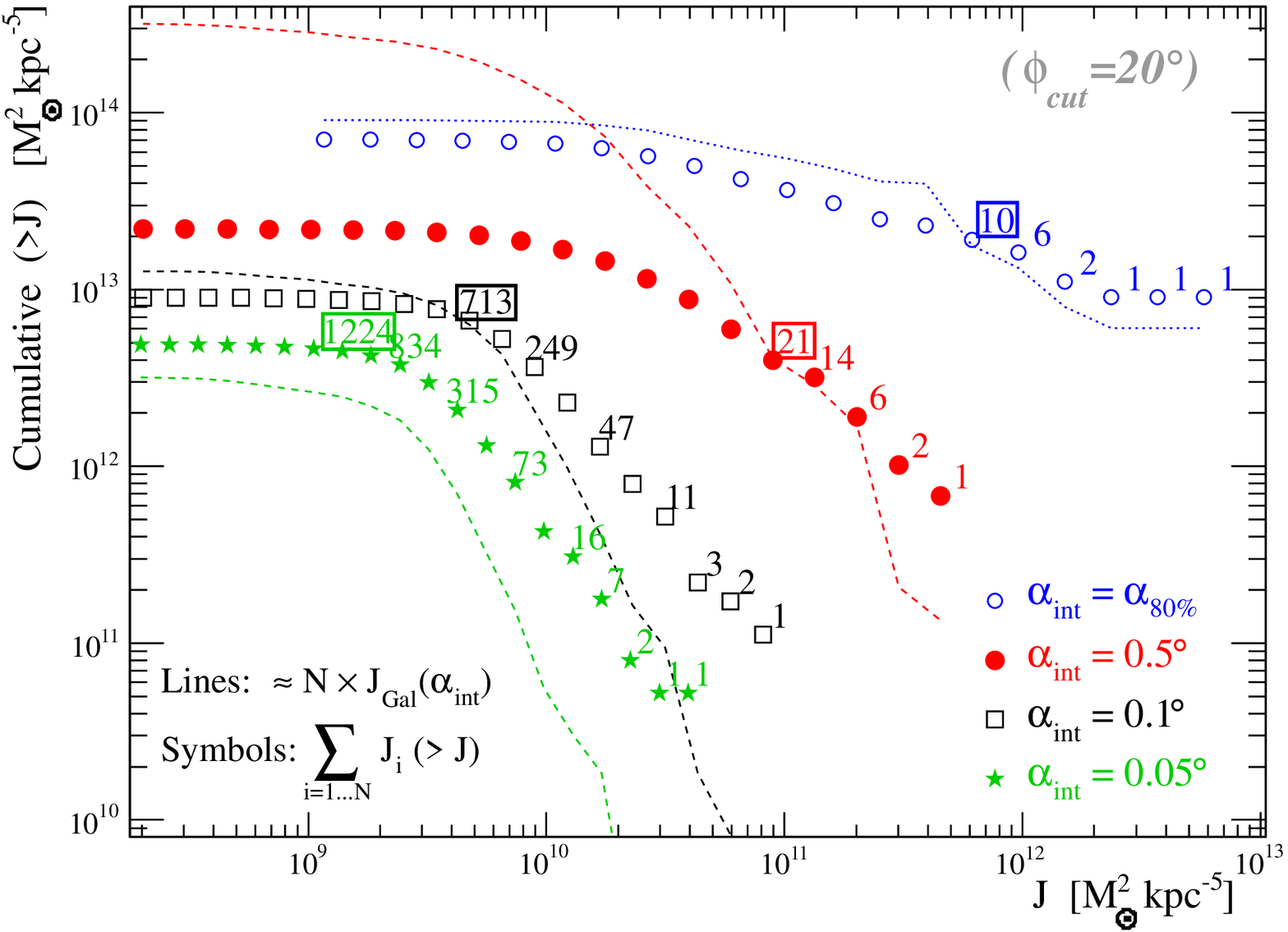}
\includegraphics[width=\linewidth]{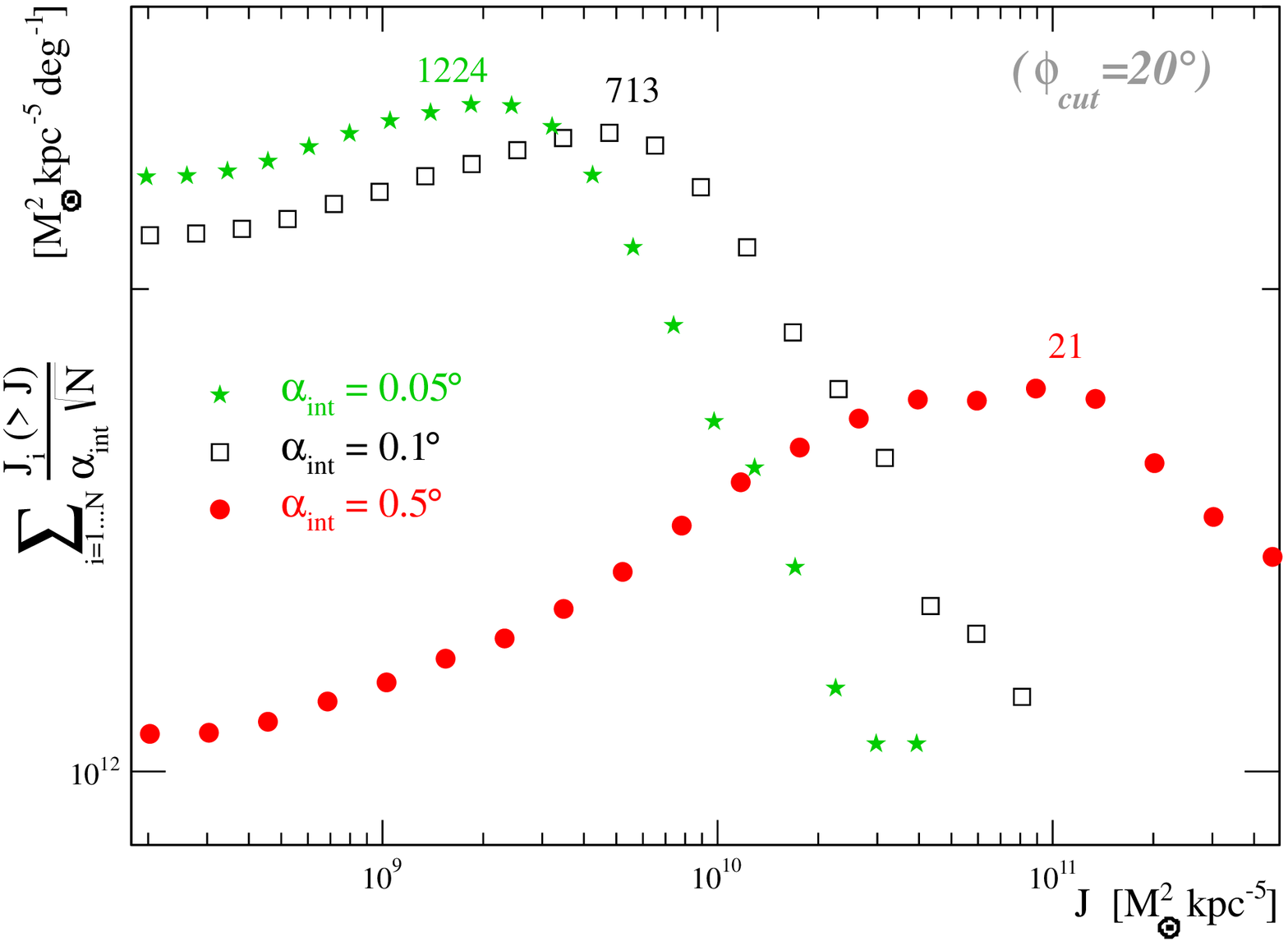}
\caption{{\bf Top:} The cumulative $J$ (i.e. $\sum_i J_i$ for all $i$ for
  which $J_i>J$). The signal and associated Galactic DM background are
  represented by an arrow and a line respectively. The open-blue circles correspond
  to an integration angle for which 80\% of the total $J$ of a galaxy cluster is
  included. Three integration angles are shown: $0.5^\circ$ in solid-red circles,
  $0.1^\circ$ in open-black squares, and $0.05^\circ$ in solid-green stars. Clusters closer than $\phi_{\rm
  cut}=20^\circ$ from the GC are discarded. {\bf Bottom:}
  $\sum_i{J_i(>J)/\alpha_{\rm int}\sqrt{N}}$ (proportional to the cumulative
  signal-to-noise for a fixed integration angle) as a function of $J$ for
  the same integration angles.}
\label{fig:integration_angle}
\end{center}
\end{figure}

\paragraph*{Signal-limited regime}
The top panel in
Fig.~\ref{fig:integration_angle} shows the cumulative distribution of $J$ for
integration angles of 0.05$^\circ$ (solid-green stars), 0.1$^\circ$ (open-black
squares), and 0.5$^\circ$ (solid-red circles) as well as $\alpha_{\rm80\%}$
(open-blue circles). The numbers denote the number of MCXC contributing to the
cumulative in a given $J$ bin. The MCXC sources are naturally ordered by $J$ in
this plot. Sources within 20$^\circ$ of the Galactic centre are excluded. The
cumulative $J_{\rm Gal}$ is also shown (dashed lines). As mentioned in
Section~\ref{sec:distribution} the {\em contrast} ($J_{\rm target}/J_{\rm Gal}$) is
related to the detectability of an object if we are only limited by the amount of
signal available. In such a regime a stacking analysis remains valid as long as we
add sources with a {\em contrast} larger than one. The boxed number in italics
indicate at what point this occurs: for an integration angle of 0.5$^\circ$ the
optimum number of objects to stack in this regime is 21. The wealth of sources in
the MCXC becomes more useful for smaller integration angles, with an optimum of
1224 objects at 0.05$^\circ$. For the latter, the {\em contrast} never falls below
one, but beyond 1224 objects, the total $J$ does not significantly increase. For
$\alpha_{\rm80\%}$ only 10 sources can be stacked before the signal is dominated by
$J_{\rm Gal}$. The total $J$ (with a {\em contrast}$>1$) available in these
scenarios is $\sum J_{\rm numbered}\approx4\times10^{12}~M_\odot$~kpc$^{-5}$ for
$\alpha_{\rm int}=0.5^\circ$, $8\times10^{12}~M_\odot$~kpc$^{-5}$ for $\alpha_{\rm
int}=0.1^\circ$, and $5\times10^{12}~M_\odot$~kpc$^{-5}$ for $\alpha_{\rm
int}=0.05^\circ$. The maximal value that can be achieved is $\sum J_{\rm
numbered}\approx2\times10^{13}~M_\odot$ if  $\alpha_{\rm int}=\alpha_{\rm80\%}$,
i.e. ten times the result that can be achieved at fixed integration angle.

\paragraph*{Background-limited regime: all-sky vs pointed instruments}

It is not just the Galactic DM background that is important in the selection of target
objects, but also the astrophysical $\gamma$-ray background. As the DM annihilation
signal is prominent at the very central part of halos, it is subject to
$\gamma$-ray and cosmic-ray contamination from astrophysical sources. Among these
are the powerful AGN (hosting a super-massive black hole) often found at the
cluster centre \citep[e.g.,][]{mcnamara07}, or intra-cluster shock-driven particle
acceleration \citep[e.g.,][]{markevich07,ensslin11,2011PhRvD..84l3509P}. This
astrophysical background will increase with the square of the integration angle.
The signal-to-noise ratio for a source is therefore proportional to
$J/\sqrt(\alpha^2)$. The cumulative signal-to-noise ratio for an all-sky instrument
(in which all objects are observed for the total observation time) is therefore
proportional to $\sum_i{J_i(>J)}/\sum{\alpha_i^2}$. For a fixed integration angle,
this is $\sum_i{J_i(>J)/\alpha_{\rm int}\sqrt{N}}$.  For an instrument that relies
on pointed observations, the amount of time spent on each source is the total
observing time available divided by the number of sources that must be observed.
Therefore, the signal-to-noise ratio is proportional to
$\sum_i{J_i(>J)}/\sum{\alpha_i^2}\sqrt{N}$.  In that case, the best strategy
appears to focus on a single bright object. As the total available
observation time is fixed, time spent observing additional sources reduces the time
spent observing the brightest target (see Section~\ref{sec:detection}).

The lower panel of Fig.~\ref{fig:integration_angle} shows
$\sum_i{J_i(>J)/\alpha_{\rm int}\sqrt{N}}$ as a function of $J$, again for
integration angles of 0.05$^\circ$ (solid-green stars), 0.1$^\circ$ (open-black
squares), and 0.5$^\circ$ (solid-red circles). The peak in these `signal-to-noise'
curves indicates the optimum number of sources to stack in the background-limited
regime, and are highlighted as 1224, 713 and 21 for 0.05$^\circ$, 0.1$^\circ$ and
0.5$^\circ$ respectively. In this plot, sources are ordered by increasing $J$-values, and therefore
only `signal-to-noise' curves can be included for fixed integration angles. For
variable integration angles, such as $\alpha_{\rm80\%}$, the signal-to-noise ratio
of each source in the catalogue will depend on the integration angle as well as $J$,
and therefore the stack must be ordered by $J/\alpha_{\rm80\%}$. In this case the
optimum number of sources is close to the full stack size, though we will see in
the following section that these optimum values change drastically when the angular
response of the instrument is considered. Examining the list in detail, it is
apparent that when ordering by $J/\alpha_{\rm80\%}$ rather than $J$, only a few
sources high-up the list swap places. The sources falling somewhere in the `top'
20-30 remain consistent.

The conclusions drawn from Fig.~\ref{fig:integration_angle} are only
valid for a instrument with a perfect angular response. In reality, the
angular response of an instrument|typically characterised by the
point spread function (PSF) which we take here to mean the 68\%
containment radius|must be combined with the integration angle in
quadrature before considering the amount of background contamination
in an observation.
In deciding which integration angle to use, we consider that, for a
small fixed angle, the cumulative $J$ is reduced since some signal
from angularly-large sources is neglected. For a large fixed angle
(e.g., 0.1$^\circ$), the cumulative $J$ increases slowly, implying that
angularly-large sources are also bright, and located near the top of
the list. Further down the list, where sources are angularly small,
large amounts of galactic contamination and astrophysical background
are included unnecessarily. Therefore a different
integration angle for each source, such as $\alpha_{\rm80\%}$, may be
optimum, and is used in the remainder of the analysis.

\subsection{Strategy for a `real' (PSF-limited) instrument}

\begin{figure}
\begin{center}
\includegraphics[width=\linewidth]{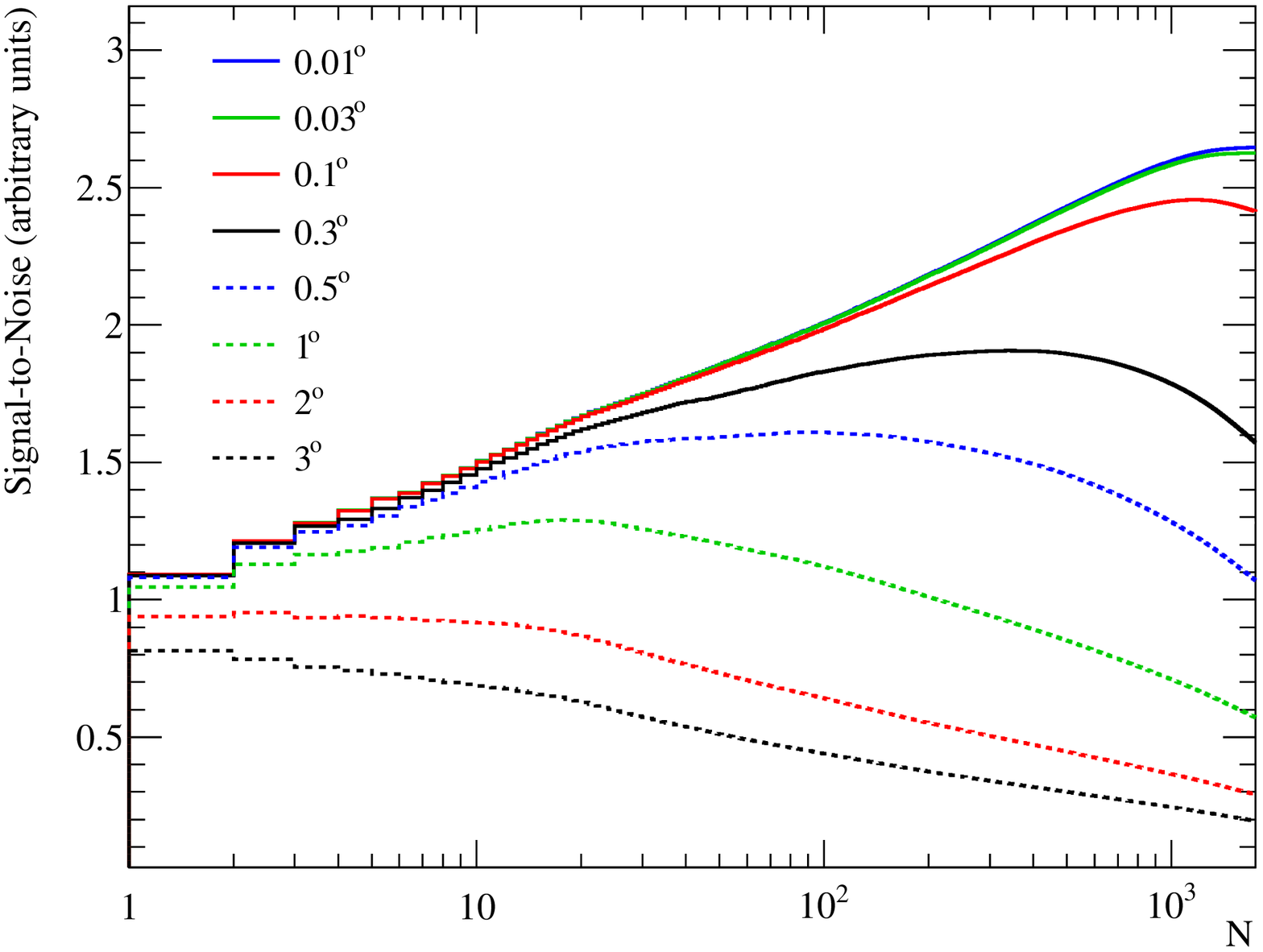}
\includegraphics[width=\linewidth]{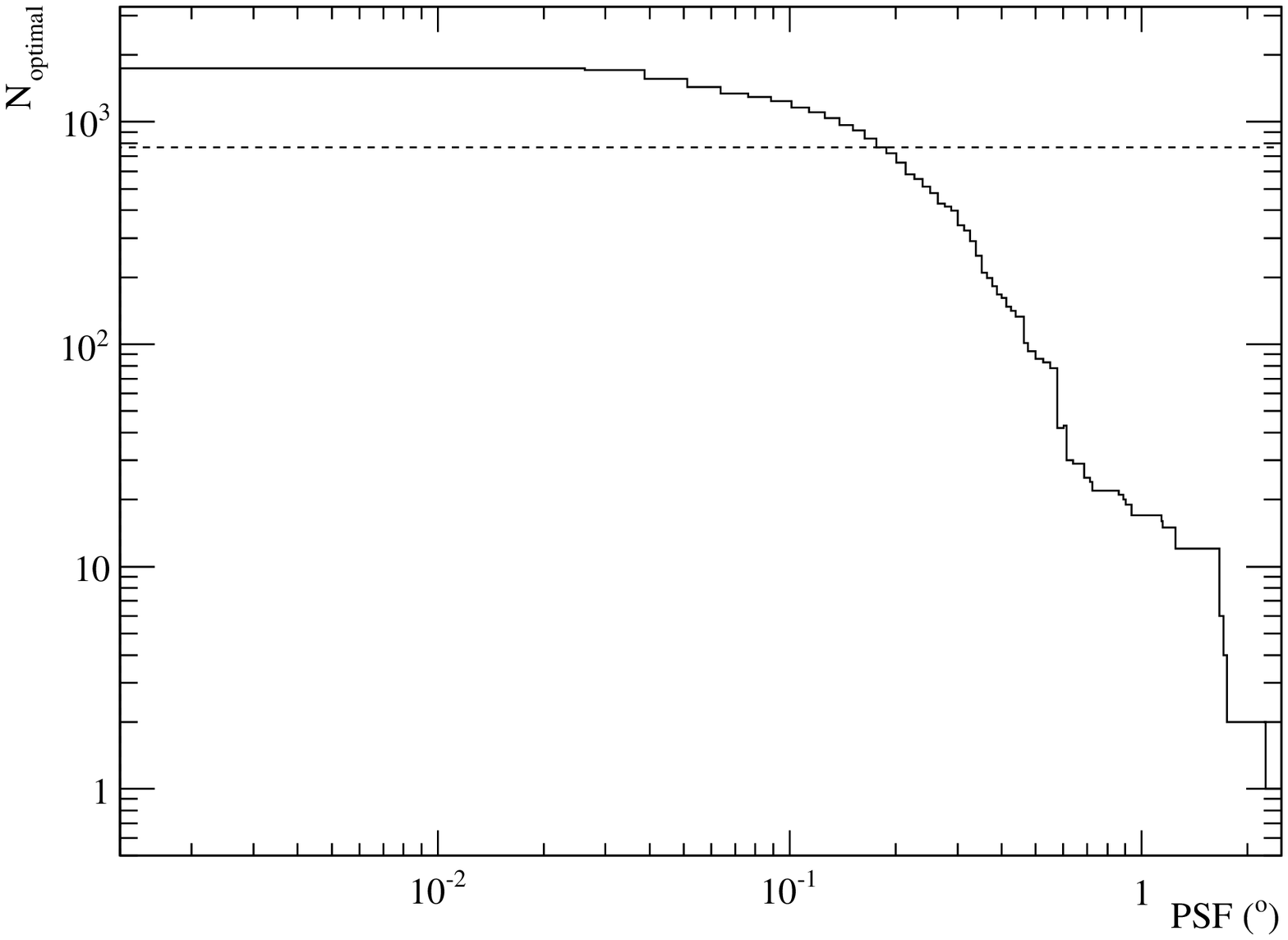}
\caption{{\bf Top panel:} The signal-to-noise ratio as a function of the 
  number of sources to stack for different values of an instrument point-spread
  function (PSF). {\bf Bottom panel:} Optimum number of sources as a function of
  the PSF for $\alpha_{\rm int}$ set to $\alpha_{80\%}$. 
  For a fixed integration angle $\alpha_{\rm int}=0.1^\circ$, this number is constant
  with PSF (dashed line). 
}
\label{fig:StoN}
\end{center}
\end{figure}

The upper panel of Fig.~\ref{fig:StoN} shows the cumulative signal-to-noise
ratio as a function of the number of sources stacked for
different values of the PSF. As the PSF worsens from 0.01$^\circ$ to
3$^\circ$, the relative signal-to-noise ratio drops and the peak
position shifts towards a smaller stack size. The peak position
indicates the optimum number of sources to stack, and is shown in the
lower panel of Fig.~\ref{fig:StoN} as a function of PSF for an
all-sky instrument. For a fixed integration angle of 0.1$^\circ$
(dashed line), the optimal number is constant with the PSF.  When
$\alpha_{\rm80\%}$ is considered, the optimal number of sources drops
as the PSF of the instrument increases. For a PSF of 0.1$^\circ$, 1200
sources should be stacked. For a PSF of 0.5$^\circ$, 90 sources should
be stacked and for a PSF of 1$^\circ$, 17 sources should be
stacked. When the PSF increases above $\sim$2$^\circ$, stacking is no
longer a valid approach, and only the brightest source should be
considered. It is not only the number of sources that should be
stacked that changes with PSF, but also the order of those
sources. Independent of the PSF the top two sources are Virgo and then
A\,426. At a PSF of $>$1$^\circ$ the third brightest source is
NGC\,4636. However, below a PSF of 1$^\circ$, A\,3526 moves into third
place. The top ten sources always contains Coma, but Fornax falls out
of the top ten when the PSF drops below $\sim$0.1$^\circ$.

\subsection{Detection Prospects}
\label{sec:detection}

In this section, we assess the DM detection prospects for the stacking
of sources from the MCXC for the Fermi-LAT all-sky $\gamma$-ray satellite,
and the envisaged array of Imaging Atmospheric Cherenkov Technique
CTA (Cherenkov Telescope Array). Whilst the design of CTA is
still evolving, performance curves for several configurations have
been released. Here, we use the so-called array layout `E', which is
described in \citet{cta:concept}.  For the Fermi-LAT, the 1-year
point-source performance curves for a high-latitude source are used
(\citealt{2009arXiv0907.0626R}). The diffuse galactic and extra-galactic
background models given by the template files \texttt{gal\_2yearp7v6\_v0.fits} and \texttt{iso\_p7v6source.txt}
respectively, which are available from the Fermi-LAT data server, are used
to obtain the background within the integration angle for each source
position on the sky.  A toy likelihood-based model, as used in
\citet{2011MNRAS.418.1526C}, is used to obtain the sensitivity of
these instruments to the DM galaxy cluster signal.

\begin{figure}
\begin{center}
\includegraphics[width=\linewidth]{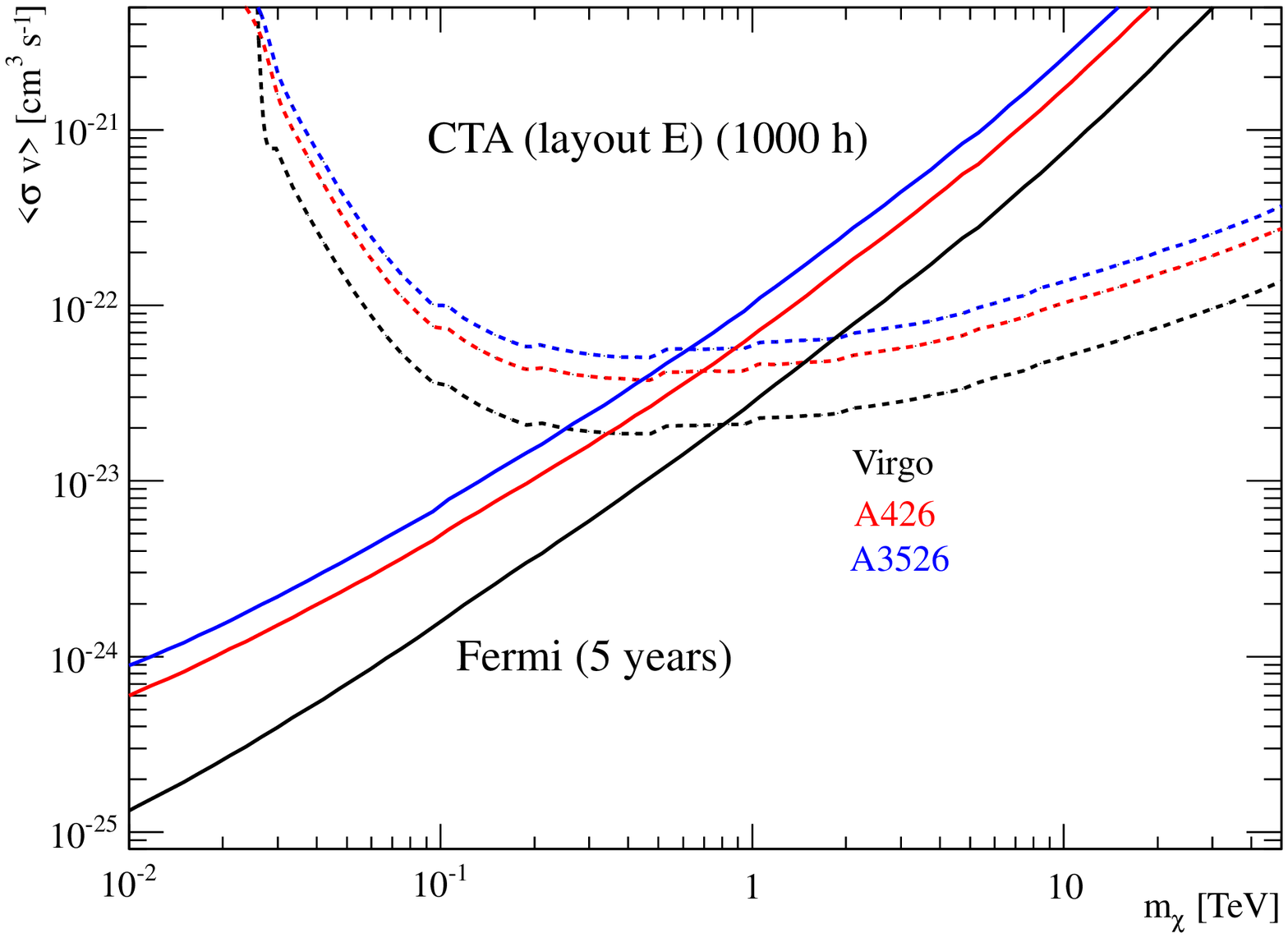}
\includegraphics[width=\linewidth]{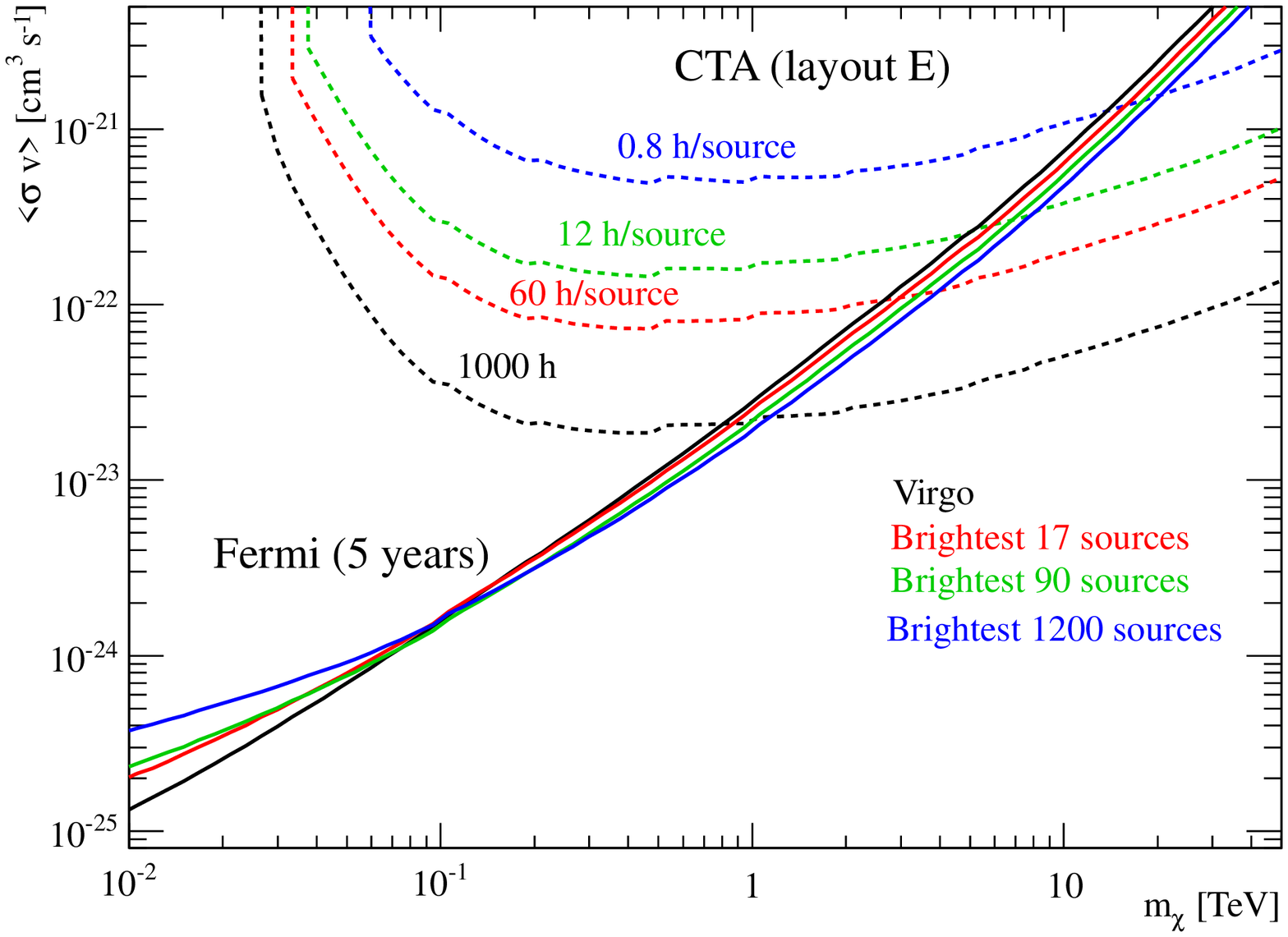}
\caption{{\bf Top panel: } the 5$\sigma$ sensitivity of Fermi-LAT (5 years
  exposure) (solid curves) and CTA (1000 hours exposure per source)
  (dashed curves) to the three brightest MCXC sources in the context
  of this work: Virgo (black), A\,426 (red) and A\,3526 (blue) when
  considering an integration angle of $\alpha_{\rm80\%}$ .  {\bf
    Bottom panel:} as above, for stack sizes of the optimum number of
  sources for a 0.1$^\circ$ (1200) (blue), 0.5$^\circ$ (90) (green)
  and 1$^\circ$ (17) (red) PSF obtained from
  Fig.~\ref{fig:StoN}. Virgo alone is again shown in black. For CTA
  the 1000 hour exposure is divided equally over the number of sources
  in the stack.}
\label{fig:sigmav}
\end{center}
\end{figure}
The top panel of Fig.~\ref{fig:sigmav} shows the sensitivity of
Fermi-LAT and CTA to the three sources from the MCXC that result in the
highest $J$ and $J/\alpha_{\rm80\%}$ (`signal-to-noise') in the
context of this work for PSFs smaller than 1$^\circ$: Virgo, A\,426,
and A\,3526 when considering an integration angle of
$\alpha_{\rm80\%}$. All curves represent a 5-sigma significance. The
Fermi-LAT curves are computed for 5 years exposure, whilst the CTA curves
assume 1000 hours observation of each source. Whilst it is unrealistic
to expect pointed observations on all three of these sources for this
duration, an equal exposure is useful in comparing the potential
targets. Virgo dominates the sensitivity for both detectors. Individual curves
were produced for the ten brightest sources, and it was found that the
top three sources shown here provide the best individual
sensitivity.

A spectrum of photon energies is associated with each DM mass. Most
sensitivity is contributed by the photon energy range close to the
peak in $E^{2}dN/dE$, which lies one order of magnitude below the DM mass 
for our assumed annihilation spectrum. Very low energy photons
(several orders of magnitude below the DM mass) contribute little to
the sensitivity due to the relatively hard signal spectrum and
overwhelming background. In our analysis we exclude photons with
energies less than 1/200 the DM mass (providing this cut lies below 10
GeV). We consider this to be a realistic approach in practice to
avoid source confusion problems due to the very poor PSF
of Fermi-LAT close to threshold. At 100~MeV for example, the Fermi-LAT PSF
is some 6$^\circ$ in radius, a region
that is likely to include several additional Fermi sources.

The lower panel of Fig.~\ref{fig:sigmav} shows the sensitivity of
Fermi-LAT and CTA when stacking the optimum number of sources determined
from the lower panel of Fig.~\ref{fig:StoN} for PSFs of: 0.1$^\circ$
(1200), 0.5$^\circ$ (90) and 1$^\circ$ (17). The brightest source,
Virgo, is shown individually. Again, the Fermi exposure is taken as 5
years. As Fermi is an all-sky instrument, each source in the stack
receives this exposure regardless of the stack size.

At DM masses below $\sim$100~GeV, the majority of photons are
collected in an energy range where the Fermi-LAT PSF is worse than a
degree. Here, the analysis falls into the background-limited
regime. Therefore stacking does not help, and just adds background,
making the sensitivity worse than Virgo alone, for example:
$\sim$3.5$\times$10$^{-25}$~cm$^3$~s$^{-1}$ to
$\sim$2.5$\times$10$^{-25}$~cm$^3$~s$^{-1}$ respectively at
$\sim$2~GeV. Searching for WIMP masses above $\sim$100 GeV, photons
begin to be included that are seen by Fermi-LAT with a better PSF. In this
mass regime, the amount of signal collected becomes important, and the
stacking helps. The analysis eventually becomes signal-limited, and
stacking improves the sensitivity by a factor of up to 1.7, from
$\sim$3$\times$10$^{-23}$~cm$^3$~s$^{-1}$ to
$\sim$1.8$\times$10$^{-23}$~cm$^3$~s$^{-1}$ at $\sim$1~TeV. This is
roughly equivalent to the improvement in signal-to-noise ratio shown
in the upper panel of Fig.~\ref{fig:sigmav} for a PSF representative
of the energy range in question. For example, at a mass of 2 TeV,
photons are included down to 10 GeV, corresponding to a PSF always
better than 0.25$^\circ$. Even at masses where an improvement with
stacking is found, beyond a stack size of 17 sources the improvement
is negligible. This is simply because the instrument PSF varies with
energy and therefore taking the optimum number of sources for a fixed
PSF is only an approximation.

In the case of CTA, we assume that a total exposure of 1000 hours is
available, and since CTA requires pointed observations, this is
reduced to $\sim$60 hours per source when 17 objects are stacked,
$\sim$12 hours per source when 90 objects are stacked, and $\sim$0.8
hours per source when 1200 objects are stacked. This effect dominates
any gain in sensitivity due to stacking, and confirms the finding of
the previous section that for an instrument requiring pointed
observations, only the brightest source should be targeted.  Note that
systematic effects are not included, and will limit the accuracy of a
1000 hour observation.

\section{Discussion}
\label{sec:discussion}

A stacking analysis of galaxy clusters may provide better limits for
indirect detection of DM than the analysis of any single object, at
least for all-sky instruments. However, this improvement is likely to
be modest for the case of annihilating dark matter. Stacking is more
promising in the case of decaying dark matter \citep{2012PhRvD..85f3517C}.
For instruments requiring pointed observations such as CTA, observing
the most promising source until the observation is systematics limited
and then moving to additional sources is a reasonable strategy. Such
an approach also mitigates against the uncertainty in the properties
of individual halos.

Limits placed on the velocity-averaged cross section depend on the
determination of $J$ not only for studies relying on known detector
sensitivities (such as this work), but also for works making use of real
data, e.g. the Fornax observation by H.E.S.S.
\citep{2012ApJ...750..123A}. We checked that given the same $J$ for a
given source, we obtain a very similar sensitivity to that estimated
in previous studies (see Appendix~\ref{app2}). In our analysis, Virgo
has the highest astrophysical factor ($J$) and best signal to noise ratio,
followed by A\,426. Several authors have suggested (based on
cluster properties given by the HIFLUGCS catalogue) that Fornax is the most
promising galaxy cluster for DM annihilation. However, as discussed above, the
MCXC provides homogenised values for $M_{500}$ based on a more accurate
gas density prescription that typically results in lower $J$ for the brightest
clusters (but note that there is no systematic trend when
all galaxy clusters are compared, see \citealt{piffaretti11}).
The differences between these two catalogues are
large enough to significantly change the conclusions of studies on the
sensitivity of current and future instruments to DM annihilation, for
example the detectability (or not) of DM  with the annihilation
cross-section expected for a thermal relic in this class of objects.
In that respect, the ranking we provide from the MCXC catalogue should
be robust, although the $J$ values calculated in this paper may still
change depending on the level of clumpiness, exact mass-concentration
relation, etc.

For all-sky instruments and in particular for Fermi-LAT, the improvement in
sensitivity obtained by stacking is at best a factor 1.7: MCXC sources with
the 1200 largest values of $J$ or $J/\alpha_{\rm80\%}$ should be
included to obtain this improvement. Additional sources do not improve
the sensitivity, as further background is integrated without
significant additional signal. This implies that the benefits of
stacking are limited by the PSF of the available all-sky $\gamma$-ray
instruments. Indeed, the PSF of Fermi-LAT at low energy is several degrees,
while the majority of MCXC targets are distant and hence subtend small angles,
with a typical $\alpha_{\rm80\%}$ of $\sim0.15^\circ$ (when substructures are
considered): an all-sky instrument with a PSF approaching
$\alpha_{\rm80\%}$ at all energies would benefit from the stacking of
all sources in the MCXC. In this case, sensitivity would then be limited only
by the available signal, and an extended catalogue|as should be provided
in a few years from now by the eROSITA mission \citep{2011SPIE.8145E.247P}|
including even fainter objects would be needed to reach a cumulative $J\sim 10^{11}-
10^{12}$.

A stack of the top 1200 objects excluding Virgo results in a
sensitivity only $\sim$15\% worse than the same stack size including
Virgo. In this case, the improvement in sensitivity between the
brightest source alone (A\,426) and the stack of 1200 objects is
nearly a factor of 3 above masses of 100 GeV. The advantage is that
the large number of clusters stacked is expected to wash out individual
uncertainties on the halo properties (e.g., the dispersion of mass-concentration
relationship). One viable strategy
might therefore be to use Virgo as an independent confirmation of the
signal established through the stacking of other clusters. Virgo
contains the known $\gamma$-ray emitter M\,87
\citep{2004NewAR..48..407B,2009ApJ...707...55A}. The
$\alpha_{\rm80\%}$ of Virgo is $\sim$0.3$^\circ$ for a smooth halo,
comparable to the Fermi-LAT PSF at the highest photon energies, but
$\sim$3$^\circ$ when substructure is considered. Disentangling the
point-like emission from M\,87 from any extended DM emission may
therefore be possible. Very recently, \cite{2012arXiv1201.1003H} have
claimed evidence at the $\sim$4$\sigma$ level for diffuse DM-like
emission from Virgo: they use photon energies detected by Fermi-LAT above
100~MeV and a full likelihood fit to a template vs. a point source.
Further Fermi-LAT observations, and deeper investigation of possible
astrophysical origins for the apparent extended emission, are required
to confirm or refute this intriguing result.

The great advantage of an all-sky instrument such as the Fermi-LAT is
the simultaneous observation of all sources. Analysis of the potential
DM signal from galaxy clusters can therefore be performed for
different numbers of stacked objects with different orderings
simultaneously. In the event of any detection from a stacked analysis,
a re-analysis on a different, more numerous, set of objects may help
to confirm the result. CTA only becomes competitive with Fermi for DM
masses above $\sim$1~TeV. However, at these energies CTA will have an
angular resolution approaching 0.02$^\circ$ and may therefore help in
isolating point-like sources from clusters (Virgo may not be the
only galaxy cluster with a $\gamma$-ray emitting source embedded within),
to aid in the choice of sources to stack for a Fermi analysis, or in a
hopeful case to rule out a point-like emitter as the source of Fermi
detection. CTA may also be critical to measure the cut-off in the DM
annihilation spectrum for heavy dark matter, and hence measure the DM
mass and establish the universality of the annihilation spectrum.

Data analysis can be optimised by adapting the integration region for
each cluster, as we have shown with the example of
$\alpha_{\rm80\%}$. We provide the necessary ingredients to refine
the analysis presented here in Appendix~\ref{app1}. From the dark
matter modelling side, a systematic study remains to be done to take
into account various DM profiles, substructure characteristics, the
mass-concentration dispersion, etc. This will be carried out in a
future work.  We reiterate here that our limit on $\langle\sigma
v\rangle$ could be changed by taking other configuration of the
substructure distribution. In the most favourable case, it would
allow to reach the benchmark value $\langle\sigma v\rangle \sim 3\,
10^{-26}$~cm$^3$~s$^{-1}$ coming from cosmological constraints.

\section*{Acknowledgements}
We thank C. Adami, S. Bryan, N. Fornengo, E. Jullo, J.-P. Kneib,
and M. Limousin  for providing us with useful references and for
fruitful discussions.
R.~W. acknowledges support from an STFC Postdoctoral Fellowship. 
%
%
\appendix
\section{The Relationship between $J$ and $\alpha_{\rm int}$}
\label{app1}

There exists a simple parametrisation to calculate $J(\alpha_{\rm int})$ for
any $\alpha_{\rm int}$, given the DM profile \citep{2012PhRvD..85f3517C}.
Indeed, we can assume that all galaxy clusters share the same DM profile. Given the mass
range span by the MCXC, we can approximate at first order their
concentration parameter to be the same. For a NFW profile,
$c(M) = R_{\rm vir}/r_s$ and we take $c(10^{14} M_\odot)\sim5$
\citep{2008MNRAS.390L..64D}. Defining
\begin{eqnarray}
  \alpha_s    &\equiv& \tan^{-1}\left(\frac{r_s}{d}\right),
     \,\,\, \alpha_{\rm max}   \equiv \tan^{-1}\left(\frac{5r_s}{d}\right),\\
  \label{eq:alphas}
  x           &\equiv& \frac{\alpha_{\rm int}}{\alpha_s}, \,\,\,
  {\rm ~~and~~~} x_{\rm max} \equiv \frac{\alpha_{\rm max}}{\alpha_s}\approx 5,
  \label{eq:xmax}
\end{eqnarray}
there is a universal dependence of the fraction of the smooth and substructure
contributions \citep{2012arXiv1203.1166M},
\[
{\cal F}_{J}(x) \equiv \frac{J(x\cdot\alpha_s)}{J_{\rm max}},
\]
which we parametrise to be (valid only for a NFW)
\begin{equation}
 {\cal F}_{\rm smooth}(x) \!\!=\!\!
    \begin{cases}
      3 x^{0.93} \text{~~~if $x \leq 10^{-2}$,}\\
      1   ~~~~~~~~~~\text{~~~if $x \geq 5$,}\\
      e^{\left[-0.086 + 0.17  \ln(x)
           - 0.092 \ln^2(x) + 0.011 \ln^3(x)\right]};\\
    \end{cases}
\end{equation}
and
\begin{equation}
 {\cal F}_{\rm subs}(x) \!\!=\!\!
    \begin{cases}
      1   ~~~~~~~~~~\text{~~~if $x \geq 5$,}\\
      e^{\left[-1.17 + 1.06 \ln(x)
           - 0.17 \ln^2(x) -0.015 \ln^3(x)\right]}.\\
    \end{cases}
\end{equation}

The `signal' $J$ can then be calculated for any integration angle, using
\begin{eqnarray}
 J_{\rm tot} (\alpha_{\rm int})&=&
      J_{\rm smooth}(0.1^\circ) \times
   \frac{{\cal F}_{\rm sm} \left(\alpha_{\rm int}/\alpha_s\right)}{
   {\cal F}_{\rm sm} \left(0.1^\circ/\alpha_s\right)}\nonumber\\
  &+&  J_{\rm subs}(0.1^\circ) \times
   \frac{{\cal F}_{\rm subs} \left(\alpha_{\rm int}/\alpha_s\right)}{
   {\cal F}_{\rm subs} \left(0.1^\circ/\alpha_s\right)}\,,
   \label{eq:F_annihl}
\end{eqnarray}
Hence, as shown in \citet{2012arXiv1203.1166M}, for DM annihilation, one needs
three quantities (available for all clusters in the Supplementary
Material---ASCII file---submitted with the paper, short sample in appendix~\ref{app3}), i.e., $\alpha_s$,
$J_{\rm smooth}(0.1^\circ)$ and $J_{\rm subs}(0.1^\circ)$.

This parametrisation describing the fraction of the signal in a given angular
region is valid down to ${\cal F}_J=10^{-3}$.

\section{A comparison of the values of $J$ obtained here to other work}
\label{app2}
DM annihilation in galaxy clusters has been studied in several papers including
\citep{2009PhRvD..80b3005J,2010JCAP...05..025A,2011JCAP...12..011S,
2011PhRvD..84l3509P,2012MNRAS.419.1721G,2012JCAP...01..042H,2012arXiv1201.0753A,2012arXiv1201.1003H}.
Below, in Table~\ref{tab:comparison}, we provide a comparison with
some of these studies, whenever the $J$ factor was available.

The calculations of the present work are consistent with those of
\citet{2011JCAP...12..011S}, \citet{2011PhRvD..84l3509P} and \citealt{2012MNRAS.419.1721G}. 
Our results for the boost values are also in agreement. Our $J$ values are also
broadly consistent though systematically lower (resp. systematically larger)
than those of \citet{2010JCAP...05..025A} and \citet{2012JCAP...01..042H}
without (resp. with) the substructure contribution. In any case, the
uncertainties quoted in these two papers (third line in the table) is probably
underestimated. Note that all the three studies rely on the HIFLUGCS catalogue
based on ROSAT and ASCA X-ray observations \citep{2007A&A...466..805C}.  The
main difference is for Fornax, which is a factor of ten larger (though the
difference is less significant if we compare with  \citealt{2011JCAP...12..011S}
results). This is due to the lower mass we infer for this cluster from the MCXC
$M_{500}$, and $R_{500}$ values, which are based on a better modelling
of the gas in the cluster \citep{piffaretti11}.

\begin{table}
  \caption{Comparison with $J$ values from other works for $\alpha_{\rm int} = 0.1^\circ$
  and $1^\circ$ respectively.}
  \label{tab:comparison}
  \centering
  \begin{tabular}{lcccccc}
  \hline
              &   \multicolumn{3}{c}{$\log_{10}\left(\frac{J(1^\circ)}{{\rm GeV}^2~{\rm cm}^{-5}}\right)$} &~~~~~& \multicolumn{2}{c}{$\log_{10}\left(\frac{J(0.1^\circ)}{{\rm GeV}^2~{\rm cm}^{-5}}\right)$} \\
Ref.          &  [1]$^\ddagger\!\!$ &    [2]$^\S\!\!$     &$\!\!\!\!$ This work$\!\!\!\!$&~~~~~&[3]$^\P\!\!$&$\!\!\!\!\!\!\!\!\!\!\!$ This work$\!\!\!\!\!\!\!\!$\\
    Error     &$\!\lesssim0.1\!$&$\!\lesssim0.2\!$&   (wo/w subs)    &~~~~~&   -   &  (no subs)  \\
  \hline
    Fornax    &      17.8       &       17.9      &  16.9~~~18.8  &~~~~~&  17.0  & 16.7 \\
    Coma      &      17.2       &       17.1      &  16.9~~~18.4  &~~~~~&  16.8  & 16.7 \\
    A1367     &        -        &       17.1      &  16.7~~~18.3  &~~~~~&   -    & 16.5 \\
    A1060     &        -        &       17.3      &  16.8~~~18.3  &~~~~~&   -    & 16.7 \\
    AWM7      &      17.1       &       17.2      &  16.8~~~18.2  &~~~~~&   -    & 16.6 \\
    NGC4636   &      17.6       &       17.5      &  17.2~~~18.2  &~~~~~&   -    & 16.9 \\
    NGC5813   &        -        &       17.3      &  17.1~~~18.1  &~~~~~&  16.4  & 16.8 \\
A3526$^\star$ &      17.4       &         -       &  17.1~~~18.1  &~~~~~&   -    & 16.9 \\
A426$^\dagger$&        -        &         -       &  17.2~~~18.1  &~~~~~&  16.9  & 17.0 \\
   Ophiuchus   &        -        &         -       &  16.8~~~18.1  &~~~~~&  16.8  & 16.7 \\
    Virgo     &        -        &         -       &  17.9~~~18.0  &~~~~~&  17.5  & 17.5 \\
    NGC5846   &        -        &         -       &  16.7~~~17.9  &~~~~~&  16.5  & 16.5 \\
  \hline
  \end{tabular}
\medskip
{\footnotesize$\!^\ddagger\!\!$\citet{2010JCAP...05..025A},$^\S\!$\citet{2012JCAP...01..042H},
$\!\!^\P\!$\citet{2011JCAP...12..011S}

$^\star$Centaurus, $^\dagger$Perseus}
\end{table}

\section{ $J$ table of the MCXC catalog}
\label{app3}

\begin{table*}
  \caption{J values  of the five first objects of the MCXC catalog. The full table is available in the online version.}
  \label{tab:JMCXC}
  \centering
  \begin{tabular}{cccccccccc}
  \hline

 NAME & Indice MCXC &  $ l$ &    $   b $ &    $  d $ &   $ \alpha_s$ &
 $ J_{{\rm sm}}$(0.1deg) & $J_{{\rm sub}}$(0.1deg) & $\alpha_{80\%}$ & $ J_{80}(\alpha_{80\%})$\\
         -  &       - &     [deg] &   [deg] &   [kpc] &    [deg] &
         $[M_{\odot}^2{\rm kpc}^{-5}]$& $[M_{\odot}^2{\rm kpc}^{-5}]$ &   [deg]  &
         $[M_{\odot}^2{\rm kpc}^{-5}] $ \\

  \hline

        UGC12890 &    1   &  101.78 & -52.48 & 1.62e+05  & 8.36e-02 &  9.69e+08  &  8.05e+09  &    2.73e-01 &     1.94e+10\\
 RXCJ0000\_4\_0237 &   2  &   94.27  & -62.62  & 1.55e+05 & 6.23e-02 &
 5.31e+08  &  5.32e+09  &    2.13e-01  &    1.01e+10 \\
 RXCJ0001\_6\_1540 &    3  &   75.13 &  -73.73  & 4.61e+05 & 4.01e-02 &  2.82e+08  &  3.35e+09  &    1.27e-01  &    4.30e+09\\
           A2692   &  4  &   104.31 & -49.00  & 6.89e+05  & 3.19e-02 &   2.16e+08  &  2.41e+09  &    1.00e-01    &  2.67e+09 \\
           A2697   &  5  &   92.17  &  -66.03  & 7.63e+05  & 3.77e-02 &  3.30e+08  &  3.19e+09  &    1.14e-01    &  3.88e+09\\

  \hline

\end{tabular}

\end{table*}

In Table~\ref{tab:JMCXC}, the $J$ values of the five first objects of
the MCXC catalog are given. \\We provide in the online version the full
table for the 1743 MCXC catalog objects.

Table~\ref{tab:JMCXC} column description :\\
NAME: Cluster name (from MCXC)\\
Indice MCXC: Indice of the cluster row in the MCXC catalogue (see \citep{piffaretti11}).\\
$l$: Galactic longitude in degree (from MCXC).\\
$b$: Galactic latitude in degree (from MCXC).\\
$d$: Angular diameter distance in kpc (from MCXC).\\
$\alpha_s$: $\arctan{(rs/d)}$ in degree (rs is the scale radius of the cluster).\\
$ J_{{\rm sm}}$(0.1deg): Astrophysical annihilation term (from smooth
halo) in  $[M_{\odot}^2{\rm kpc}^{-5}]$ for $\alpha_{{\rm int}}=0.1$ deg.\\
$J_{{\rm sub}}$(0.1deg): Astrophysical annihilation term (from
substructures) in  $[M_{\odot}^2{\rm kpc}^{-5}]$ for $\alpha_{{\rm int}}=0.1$ deg.\\
$\alpha_{80\%}$:  Angle in degree containing 80$\%$ of the total J.\\
$J_{80}$: Value of annihilation signal for $\alpha_{{\rm int}}=\alpha_{80\%}$.\\

\label{lastpage}
\bibliography{cluster_annihil}

\end{document}